\documentclass[aps,prc,preprint,showpacs,showkeys,superscriptaddress]{revtex4-1}

\usepackage{amsfonts}
\usepackage{amsmath}
\usepackage{amssymb}
\usepackage{graphicx}
\graphicspath{{Content/Figs/}{Figs/}}
\usepackage{xcolor}
\usepackage[version=3]{mhchem}
\usepackage{url}
\usepackage{natbib}
\usepackage[colorlinks,urlcolor=blue,citecolor=blue]{hyperref}
\usepackage{simpler-wick}
\newcommand{\basbar}[1]{\,\overline{\!{#1}}}
\usepackage{tensor}
\usepackage{slashed}
\usepackage{tikz}
\usepackage{wrapfig}

\begin{document}
\title{Nucleon-Nucleon elastic and inelastic scattering using quantum field theory: a comparative study}
\author{Raghad Al-Bakri}
\affiliation{Physics Department, College of Science, Qassim University, Qassim, Burydah 51411, KSA. }

\author{Bassam Shehadeh}
\affiliation{Physics Department, College of Science, Qassim University, Qassim, Burydah 51411, KSA. }

\begin{abstract}
In this paper, we use Yukawa theory to calculate differential and total cross-sections for elastic and inelastic scattering in nucleon-nucleon interactions. We start from the fundamental Lagrangian and derive the $T$-matrix and hence the invariant scattering matrix leading towards the differential and total cross section. We perform calculations utilizing two types of Yukawa interaction terms: 1) scalar current, and 2) pseudoscalar current. We also carry out calculations in a laboratory frame. Calculated results using scalar current exhibits the exact features of elastic and inelastic scattering. The results agree very well with the differential cross-sections for experimental data. The pseudoscalar current calculations do not offer reasonable results and features of the NN interactions.
The simplicity of the theory makes it all the more impressive, thus it can be used in place of more complicated field theories to describe NN interactions.
\end{abstract}

\volumeyear{year}
\volumenumber{number}
\issuenumber{number}
\eid{identifier}
\date[Date text]{date}
\received[Received text]{date}
\revised[Revised text]{date}
\accepted[Accepted text]{date}
\published[Published text]{date}
\maketitle

\section{Introduction}

In this paper, we consider the Yukawa one-pion-exchange theory. This theory
was proposed by Yukawa \cite{yukawa1935} as a model for the nucleon-nucleon
(NN) interaction. The NN scattering by the Yukawa one-pion theory is carried
out using the effective interacting Lagrangian within the quantum field
theory scheme. The treatment and the calculations of dynamical variables,
such as energies and momenta are carried out using relativistic kinematics.
We calculated the NN scattering amplitude by implementing Wick's theory of
time-ordering operators \cite{peskin2018} for the interaction of the Dirac field
with the scalar field. Since pions are pseudoscalar fields, we also perform
similar calculations for the scattering amplitude considering a pseudoscalar
current for the elastic and inelastic scattering amplitude using
pseudoscalar current in the interaction term of the Lagrangian density.

Unlike using Yukawa one-pion-exchange-potential (OPEP), in this work, we
start from the Lagrangian density, where we use the interaction term of the
nucleon field (Dirac field) with the pion field (scalar Klein-Gordon field)
to obtain the $T$-matrix and compute the scattering amplitude. This way, we
guarantee to obtain the correct spin summation and mass distribution,
especially when considering inelastic scattering, where different spin
and mass particles are created, and the pseudoscalar interaction. This way
we can accurately describe the production of hadron resonances in any
half-odd spin.

Previous work with similar calculations is given in reference \cite{norbury2010}.
The paper assumes scattering amplitude consists of direct and
exchange terms of Mandelstam variables where term factors are pre-assumed 
\cite{norbury2010, yao2006}. In contrast, in this work, these terms are
derived from field theory, without using Feynman calculus, and the coupling
constant is modeled using NN potential strength. Another advantage of such
a method is to set the coupling constant as a floating parameter used to fit
the experimental data.

In this paper we show the quantum field theory of NN scattering by
exchanging one pion, then derive the scattering amplitude and the
differential cross section for both, elastic and inelastic scattering, using
scalar current and pseudoscalar current. The results are presented in the
results and discussion section. A comparison between the theory and the
experimental data is presented.

\section{Theory}

\subsection{Quantum field theory of the fermion-meson interaction}

The quantum field theory is a fundamental theory that succeeded in
describing fundamental interactions, such as electromagnetic, weak, and
strong interactions. These interactions are thought of as exchanges of
particles called mediators. Using this analogy, Yukawa proposed that nuclear
force results from the exchange of particles. Their masses, spin, and charge
match those of pions. Pions are spin 0 particles, needed to couple two spin-$%
\frac{1}{2}$ nucleons. The effective range of nuclear force imposes the mass
value of the pion mediators. Since nuclear force is charge independent, we
need pions to couple proton-proton, proton-neutron, and neutron-neutron.
This means three-charge states are required for the pions. There are three
different flavors of pions: $\pi ^{+}$, $\pi ^{-}$, and $\pi ^{0}$, has
triplet isospin quantum number $\tau =1$ with projection $\tau _{3}=+1,-1,0$
, respectively. Since pions are mesons, they have composite particles of
quark-antiquark, they are called pseudoscalar particles.

Fundamentally the NN interaction by exchanging pions is viewed as an
interaction of spin-$\frac{1}{2}$ Dirac field with spin-$0$ Klein-Gordon
field. The Lagrangian density thus consists of the Dirac Lagrangian as a
nucleon field (of mass $M$ \ and spin $\frac{1}{2}$) interacting with the
scalar field of pions, of mass $m$ and spin $0$ \cite%
{peskin2018,rpp_pdg_2022}. The interaction term of the Lagrangian is called
the Yukawa term. The Lagrangian becomes%
\begin{equation}
\mathcal{L=}\frac{1}{2}\partial _{\mu }\phi \partial ^{\mu }\phi -\frac{1}{2}%
m^{2}\phi ^{2}+\basbar{\psi }\left( i\gamma ^{\mu }\partial _{\mu
}-M\right) \psi -\lambda \phi \basbar{\psi }\psi .  \label{eq1}
\end{equation}%
Such Lagrangian does exist in the standard model to describe the interaction
between fermions, such as leptons and quarks, and Higgs bosons. Spin is
accounted for within Dirac Lagrangian, however, we omit the isospin effect.
The Feynman diagram for the Lagrangian for Yukawa theory is

\tikzset{every picture/.style={line width=0.75pt}} 
\begin{tikzpicture}[x=0.75pt,y=0.75pt,yscale=-1,xscale=1]
\draw    (50,270) -- (118,140.67) ;
\draw [shift={(84,205.33)}, rotate = 117.73] [fill={rgb, 255:red, 0; green, 0; blue, 0 }  ][line width=0.08]  [draw opacity=0] (10.72,-5.15) -- (0,0) -- (10.72,5.15) -- (7.12,0) -- cycle    ;
\draw    (278,269.33) -- (210,140) ;
\draw [shift={(244,204.67)}, rotate = 62.27] [fill={rgb, 255:red, 0; green, 0; blue, 0 }  ][line width=0.08]  [draw opacity=0] (10.72,-5.15) -- (0,0) -- (10.72,5.15) -- (7.12,0) -- cycle    ;
\draw    (210,140) -- (278,10.67) ;
\draw [shift={(244,75.33)}, rotate = 117.73] [fill={rgb, 255:red, 0; green, 0; blue, 0 }  ][line width=0.08]  [draw opacity=0] (10.72,-5.15) -- (0,0) -- (10.72,5.15) -- (7.12,0) -- cycle    ;
\draw    (118,140.67) -- (50,11.33) ;
\draw [shift={(84,76)}, rotate = 62.27] [fill={rgb, 255:red, 0; green, 0; blue, 0 }  ][line width=0.08]  [draw opacity=0] (10.72,-5.15) -- (0,0) -- (10.72,5.15) -- (7.12,0) -- cycle    ;
\draw [line width=1.5]  [dash pattern={on 5.63pt off 4.5pt}]  (118,140.67) -- (210,140) ;
\draw    (395,271) -- (463,141.67) ;
\draw [shift={(429,206.33)}, rotate = 117.73] [fill={rgb, 255:red, 0; green, 0; blue, 0 }  ][line width=0.08]  [draw opacity=0] (10.72,-5.15) -- (0,0) -- (10.72,5.15) -- (7.12,0) -- cycle    ;
\draw    (623,270.33) -- (555,141) ;
\draw [shift={(589,205.67)}, rotate = 62.27] [fill={rgb, 255:red, 0; green, 0; blue, 0 }  ][line width=0.08]  [draw opacity=0] (10.72,-5.15) -- (0,0) -- (10.72,5.15) -- (7.12,0) -- cycle    ;
\draw    (463,141.67) -- (531,12.33) ;
\draw [shift={(497,77)}, rotate = 117.73] [fill={rgb, 255:red, 0; green, 0; blue, 0 }  ][line width=0.08]  [draw opacity=0] (10.72,-5.15) -- (0,0) -- (10.72,5.15) -- (7.12,0) -- cycle    ;
\draw    (555,141) -- (487,11.67) ;
\draw [shift={(521,76.33)}, rotate = 62.27] [fill={rgb, 255:red, 0; green, 0; blue, 0 }  ][line width=0.08]  [draw opacity=0] (10.72,-5.15) -- (0,0) -- (10.72,5.15) -- (7.12,0) -- cycle    ;
\draw [line width=1.5]  [dash pattern={on 5.63pt off 4.5pt}]  (463,141.67) -- (555,141) ;
\draw (61,251.4) node [anchor=north west][inner sep=0.75pt]    {$u^{s}( p)$};
\draw (231,252.4) node [anchor=north west][inner sep=0.75pt]    {$u^{r}( q)$};
\draw (71,21.4) node [anchor=north west][inner sep=0.75pt]    {$u^{s'}( p')$};
\draw (214,22.4) node [anchor=north west][inner sep=0.75pt]    {$u^{r'}( q')$};
\draw (402,252.4) node [anchor=north west][inner sep=0.75pt]    {$u^{s}( p)$};
\draw (571,252.4) node [anchor=north west][inner sep=0.75pt]    {$u^{r}( q)$};
\draw (531,12.4) node [anchor=north west][inner sep=0.75pt]    {$u^{r'}( q')$};
\draw (441,12.4) node [anchor=north west][inner sep=0.75pt]    {$u^{s'}( p')$};
\draw (161,142.4) node [anchor=north west][inner sep=0.75pt]    {$\pi $};
\draw (506,142.4) node [anchor=north west][inner sep=0.75pt]    {$\pi $};
\end{tikzpicture}

In the case of NN scattering, the initial and final nucleon fields are%
\begin{equation*}
\psi \left( p\right) \psi \left( q\right) \longrightarrow \psi \left(
p^{\prime }\right) \psi \left( q^{\prime }\right) .
\end{equation*}%
The initial state is 
\begin{equation}
\left\vert i\right\rangle =\sqrt{2E_{\mathbf{p}}}\sqrt{2E_{\mathbf{q}}}a_{%
\mathbf{p}}^{s^{\dagger }}a_{\mathbf{q}}^{r^{\dagger }}\left\vert
0\right\rangle \equiv \left\vert \mathbf{p,}s\mathbf{;q,}r\right\rangle ,
\label{eq1_1}
\end{equation}%
whereas the final state becomes%
\begin{equation}
\left\vert f\right\rangle =\sqrt{2E_{\mathbf{p}^{\prime }}}\sqrt{2E_{\mathbf{%
q}^{\prime }}}a_{\mathbf{p}^{\prime }}^{s^{\prime \dagger }}a_{\mathbf{q}%
^{\prime }}^{r^{\prime \dagger }}\left\vert 0\right\rangle \equiv \left\vert 
\mathbf{p}^{\prime }\mathbf{,}s^{\prime }\mathbf{;q}^{\prime }\mathbf{,}%
r^{\prime }\right\rangle .  \label{eq1_2}
\end{equation}%
The final state in bra space is%
\begin{equation}
\left\langle f\right\vert =\sqrt{2E_{\mathbf{p}^{\prime }}}\sqrt{2E_{\mathbf{%
q}^{\prime }}}\left\langle 0\right\vert a_{\mathbf{q}^{\prime }}^{r^{\prime
}}a_{\mathbf{p}^{\prime }}^{s^{\prime }}.  \label{eq1_3}
\end{equation}%
For the $1^{st}$ order Feynman diagram two vertices%
\begin{equation}
\left\langle f\right\vert S-1\left\vert i\right\rangle \propto \frac{\left(
-i\lambda \right) ^{2}}{2!}\int d^{4}xd^{4}yT\left[ \basbar{\psi }\left(
x\right) \psi \left( x\right) \phi \left( x\right) \basbar{\psi }\left(
y\right) \psi \left( y\right) \phi \left( y\right) \right] .  \label{eq2}
\end{equation}%
Where $S-1$ represents the particles that cause the interaction, which is
known as the $T$-Matrix, and defined by the invariant matrix element $%
\mathcal{M}$. We need to emphasize that all fields are in interaction
picture. All contractions are either disconnected or cancel out each other
except for for%
\begin{eqnarray*}
T\left[ \basbar{\psi }\left( x\right) \psi \left( x\right) \phi \left(
x\right) \basbar{\psi }\left( y\right) \psi \left( y\right) \phi \left(
y\right) \right] &=&N\left[ \basbar{\psi }\left( x\right) \psi \left(
x\right) \phi \left( x\right) \basbar{\psi }\left( y\right) \psi \left(
y\right) \phi \left( y\right) \right] , \\
&=&N\left[ \basbar{\psi }\left( x\right) \psi \left( x\right) \basbar{%
\psi }\left( y\right) \psi \left( y\right) \right] \wick{\c \phi (x) \c \phi (y)},\\
&=&N\left[ \basbar{\psi }\left( x\right) \psi \left( x\right) \basbar{%
\psi }\left( y\right) \psi \left( y\right) \right] S_{F}\left( x-y\right),
\end{eqnarray*}%
where $S_{F}\left( x-y\right) $ is the contraction of the field $\phi
(x)\phi (y)$. The effect of the normally ordered operator on the initial
state is%
\begin{equation}
N\left[ \basbar{\psi }\left( x\right) \psi \left( x\right) \basbar{\psi }%
\left( y\right) \psi \left( y\right) \right] \left\vert i\right\rangle =N%
\left[ \basbar{\psi }\left( x\right) \psi \left( x\right) \basbar{\psi }%
\left( y\right) \psi \left( y\right) \right] \sqrt{2E_{\mathbf{p}}}\sqrt{2E_{%
\mathbf{q}}}a_{\mathbf{p}}^{s^{\dagger }}a_{\mathbf{q}}^{r^{\dagger
}}\left\vert 0\right\rangle .  \label{eq3_1}
\end{equation}%
Make use of the quantized Dirac field \cite{peskin2018,weinberg1995}%
\begin{equation}
\psi (x)=\int \frac{d^{3}p}{(2\pi )^{3}}\frac{1}{\sqrt{2E_{\mathbf{p}}}}%
\sum_{s=1,2}\left( a_{\mathbf{p}}^{s}u^{s}(p)e^{-ip\cdot x}+{b_{\mathbf{p}%
}^{s}}^{\dagger }v^{s}(p)e^{ip\cdot x}\right) ;  \label{QDF13}
\end{equation}%
\begin{equation}
\bar{\psi}(x)=\int \frac{d^{3}p}{(2\pi )^{3}}\frac{1}{\sqrt{2E_{\mathbf{p}}}}%
\sum_{s=1,2}\left( b_{\mathbf{p}}^{s}\bar{v}^{s}(p)e^{-ip\cdot x}+{a_{%
\mathbf{p}}^{s}}^{\dagger }\bar{u}^{s}(p)e^{ip\cdot x}\right) ,
\label{QDF13_2}
\end{equation}%
where the creation and annihilation operators obey the anticommutation
relations \cite{peskin2018}%
\begin{equation}
\left\{ a_{\mathbf{p}}^{s},{a_{\mathbf{q}}^{r}}^{\dagger }\right\} =\left\{
b_{\mathbf{p}}^{s},{b_{\mathbf{q}}^{r}}^{\dagger }\right\} =(2\pi
)^{3}\delta ^{(3)}(\mathbf{p}-\mathbf{q})\delta ^{sr},  \label{QDF10}
\end{equation}%
with all other anticommutators equal to zero. Anticommutations (\ref{QDF10})
leads to the equal-time anticommutation relations for $\psi $ and $\psi
^{\dagger }$ to be%
\begin{equation}
\left\{ \psi _{a}(\mathbf{x}),\psi _{b}^{\dagger }(\mathbf{y})\right\}
=\delta ^{(3)}(\mathbf{x}-\mathbf{y})\delta _{ab},  \label{QDF14}
\end{equation}%
where%
\begin{equation}
\left\{ \psi _{a}(\mathbf{x}),\psi _{b}(\mathbf{y})\right\} =\left\{ \psi
_{a}^{\dagger }(\mathbf{x}),\psi _{b}^{\dagger }(\mathbf{y})\right\} =0.
\label{QDF14_2}
\end{equation}%
Since there are no anti-particle contributions to the $S$-matrix, we can
ignore the anti-fermion terms in eqs. (\ref{QDF13}) and (\ref{QDF13_2}). In
general these terms automatically cancel out due to the normal ordering
of the creations and annihilations operators. Therefore, eq.(\ref{eq3_1})
becomes%
\begin{eqnarray}
&&N\left[ \basbar{\psi }\left( x\right) \psi \left( x\right) \basbar{%
\psi }\left( y\right) \psi \left( y\right) \right] \left\vert i\right\rangle 
\notag \\
&=&\sum_{t,v}\int \frac{d^{3}k_{1}}{\left( 2\pi \right) ^{3}}\frac{1}{\sqrt{%
2E_{\mathbf{k}_{1}}}}\int \frac{d^{3}k_{2}}{\left( 2\pi \right) ^{3}}\frac{1%
}{\sqrt{2E_{\mathbf{k}_{2}}}}\times   \notag \\
&&\left[ \basbar{\psi }\left( x\right) u^{t}\left( k_{1}\right) \right]
e^{-ik_{1}\cdot x}\left[ \basbar{\psi }\left( y\right) u^{v}\left(
k_{2}\right) \right] e^{-ik_{2}\cdot y}\sqrt{2E_{\mathbf{p}}}\sqrt{2E_{%
\mathbf{q}}}a_{\mathbf{k}_{1}}^{t}a_{\mathbf{k}_{2}}^{v}a_{\mathbf{p}%
}^{s^{\dagger }}a_{\mathbf{q}}^{r^{\dagger }}\left\vert 0\right\rangle .
\label{eq4}
\end{eqnarray}%
We need to implement canonical anti-commutation of the Dirac field given in
eq.(\ref{QDF10}) to have normally ordered operator ($a^{\dagger }$'s to the
left and $a$'s to the right). Using the anti-commutation (\ref{QDF10}) we
can compute the creations and annihilations operators on the vacuum state $%
\left\vert 0\right\rangle $ to get%
\begin{equation*}
a_{\mathbf{k}_{1}}^{t}a_{\mathbf{k}_{2}}^{v}a_{\mathbf{p}}^{s^{\dagger }}a_{%
\mathbf{q}}^{r^{\dagger }}\left\vert 0\right\rangle =\left\{ 
\begin{array}{c}
(2\pi )^{6}\delta ^{(3)}(\mathbf{k}_{2}-\mathbf{p})\delta ^{(3)}(\mathbf{k}%
_{1}-\mathbf{q})\delta ^{vs}\delta ^{tr}- \\ 
(2\pi )^{6}\delta ^{(3)}(\mathbf{k}_{2}-\mathbf{q})\delta ^{(3)}(\mathbf{k}%
_{1}-\mathbf{p})\delta ^{vr}\delta ^{ts}%
\end{array}%
\right\} \left\vert 0\right\rangle .
\end{equation*}%
substitute into eq.(\ref{eq4}), after applying the summation over spin, with
the aid of the Kronecker delta and carry out the integrals with the aid of Dirac
delta's we obtain,%
\begin{eqnarray}
N\left[ \basbar{\psi }\left( x\right) \psi \left( x\right) \basbar{\psi }%
\left( y\right) \psi \left( y\right) \right] \left\vert i\right\rangle  &=&%
\left[ \basbar{\psi }\left( x\right) u^{r}\left( q\right) \right]
e^{-iq\cdot x}\left[ \basbar{\psi }\left( y\right) u^{s}\left( p\right) %
\right] e^{-ip\cdot y}\left\vert 0\right\rangle   \notag \\
&&-\left[ \basbar{\psi }\left( x\right) u^{s}\left( p\right) \right]
e^{-ip\cdot x}\left[ \basbar{\psi }\left( y\right) u^{r}\left( q\right) %
\right] e^{-iq\cdot y}\left\vert 0\right\rangle .  \label{eq5}
\end{eqnarray}%
Now, let us multiply this equation with the bra of the final state $%
\left\langle f\right\vert ,$ given in eq.(\ref{eq1_3}), the amplitude of the
normally ordered operator becomes%
\begin{eqnarray}
&&\left\langle f\right\vert N\left[ \basbar{\psi }\left( x\right) \psi
\left( x\right) \basbar{\psi }\left( y\right) \psi \left( y\right) \right]
\left\vert i\right\rangle =  \notag \\
&&\sqrt{2E_{\mathbf{p}^{\prime }}}\sqrt{2E_{\mathbf{q}^{\prime }}}%
\left\langle 0\right\vert a_{\mathbf{q}^{\prime }}^{r^{\prime }}a_{\mathbf{p}%
^{\prime }}^{s^{\prime }}\left[ \basbar{\psi }\left( x\right) u^{r}\left(
q\right) \right] e^{-iq\cdot x}\left[ \basbar{\psi }\left( y\right)
u^{s}\left( p\right) \right] e^{-ip\cdot y}\left\vert 0\right\rangle - 
\notag \\
&&\sqrt{2E_{\mathbf{p}^{\prime }}}\sqrt{2E_{\mathbf{q}^{\prime }}}%
\left\langle 0\right\vert a_{\mathbf{q}^{\prime }}^{r^{\prime }}a_{\mathbf{p}%
^{\prime }}^{s^{\prime }}\left[ \basbar{\psi }\left( x\right) u^{s}\left(
p\right) \right] e^{-ip\cdot x}\left[ \basbar{\psi }\left( y\right)
u^{r}\left( q\right) \right] e^{-iq\cdot y}\left\vert 0\right\rangle .
\label{eq6}
\end{eqnarray}%
Substitute for the quantized Dirac field in eq.(\ref{QDF13_2}), ignoring the
anti-fermion terms since they yield no contribution to the amplitude,
eq.(\ref{eq6}) becomes
\begin{eqnarray}
&&\left\langle f\right\vert N\left[ \basbar{\psi }\left( x\right) \psi
\left( x\right) \basbar{\psi }\left( y\right) \psi \left( y\right) \right]
\left\vert i\right\rangle = \notag \\
&&\left\langle 0\right\vert a_{\mathbf{q}^{\prime }}^{r^{\prime }}a_{%
\mathbf{p}^{\prime }}^{s^{\prime }}a_{\mathbf{k}_{1}^{\prime }}^{t^{\prime
\dagger }}a_{\mathbf{k}_{2}^{\prime }}^{\nu ^{\prime \dagger }}\sqrt{2E_{%
\mathbf{p}^{\prime }}}\sqrt{2E_{\mathbf{q}^{\prime }}}\sum_{t^{\prime },\nu
^{\prime }}\int \frac{d^{3}k_{1}^{\prime }}{\left( 2\pi \right) ^{3}}\frac{1%
}{\sqrt{2E_{\mathbf{k}_{1}^{\prime }}}}\int \frac{d^{3}k_{2}^{\prime }}{%
\left( 2\pi \right) ^{3}}\frac{1}{\sqrt{2E_{\mathbf{k}_{2}^{\prime }}}}%
\times   \notag \\
&&\left[ \basbar{u}^{t^{\prime }}\left( k_{1}^{\prime }\right) u^{r}\left(
q\right) \right] e^{ik_{1}^{\prime }\cdot x}e^{-iq\cdot x}\left[ \basbar{u}%
^{\nu ^{\prime }}\left( k_{2}^{\prime }\right) u^{s}\left( p\right) \right]
e^{ik_{2}^{\prime }\cdot y}e^{-ip\cdot y}\left\vert 0\right\rangle -  \notag
\\
&&\left\langle 0\right\vert a_{\mathbf{q}^{\prime }}^{r^{\prime }}a_{\mathbf{%
p}^{\prime }}^{s^{\prime }}a_{\mathbf{k}_{1}^{\prime }}^{t^{\prime \dagger
}}a_{\mathbf{k}_{2}^{\prime }}^{\nu ^{\prime \dagger }}\sqrt{2E_{\mathbf{p}%
^{\prime }}}\sqrt{2E_{\mathbf{q}^{\prime }}}\sum_{t^{\prime },\nu ^{\prime
}}\int \frac{d^{3}k_{1}^{\prime }}{\left( 2\pi \right) ^{3}}\frac{1}{\sqrt{%
2E_{\mathbf{k}_{1}^{\prime }}}}\int \frac{d^{3}k_{2}^{\prime }}{\left( 2\pi
\right) ^{3}}\frac{1}{\sqrt{2E_{\mathbf{k}_{2}^{\prime }}}}\times   \notag \\
&&\left[ \basbar{u}^{t^{\prime }}\left( k_{1}^{\prime }\right) u^{s}\left(
p\right) \right] e^{ik_{1}^{\prime }\cdot x}e^{-ip\cdot x}\left[ \basbar{u}%
^{\nu ^{\prime }}\left( k_{2}^{\prime }\right) u^{r}\left( q\right) \right]
e^{ik_{2}^{\prime }\cdot y}e^{-iq\cdot y}\left\vert 0\right\rangle ,
\label{eq6_1}
\end{eqnarray}%
Using the anti-commutation (\ref{QDF10}) we can compute the creation and
annihilation operators on the bra vacuum state $\left\langle 0\right\vert $
to get%
\begin{equation*}
\left\langle 0\right\vert a_{\mathbf{q}^{\prime }}^{r^{\prime }}a_{\mathbf{p}%
^{\prime }}^{s^{\prime }}a_{\mathbf{k}_{1}^{\prime }}^{t^{\prime \dagger
}}a_{\mathbf{k}_{2}^{\prime }}^{\nu ^{\prime \dagger }}=\left\langle
0\right\vert \left[ 
\begin{array}{c}
(2\pi )^{6}\delta ^{(3)}(\mathbf{p}^{\prime }-\mathbf{k}_{1}^{\prime
})\delta ^{(3)}(\mathbf{q}^{\prime }-\mathbf{k}_{2}^{\prime })\delta
^{s^{\prime }t^{\prime }}\delta ^{r^{\prime }\nu ^{\prime }} \\ 
-(2\pi )^{6}\delta ^{(3)}(\mathbf{q}^{\prime }-\mathbf{k}_{1}^{\prime
})\delta ^{(3)}(\mathbf{p}^{\prime }-\mathbf{k}_{2}^{\prime })\delta
^{r^{\prime }t^{\prime }}\delta ^{s^{\prime }\nu ^{\prime }}%
\end{array}%
\right] .
\end{equation*}%
Substitute into eq.(\ref{eq6_1}), after applying the summation over spin,
with the aid of the Kronecker delta and carry out the integrals with the aid of
the Dirac delta we find that the amplitude of the normally ordered operator
becomes in this form%
\begin{eqnarray}
\left\langle f\right\vert N\left[ \basbar{\psi }\left( x\right) \psi
\left( x\right) \basbar{\psi }\left( y\right) \psi \left( y\right) \right]
\left\vert i\right\rangle &=&\left[ \basbar{u}^{s^{\prime }}\left(
p^{\prime }\right) u^{r}\left( q\right) \right] \left[ \basbar{u}%
^{r^{\prime }}\left( q^{\prime }\right) u^{s}\left( p\right) \right]
e^{i\cdot x\left( p^{\prime }-q\right) }e^{i\cdot y\left( q^{\prime
}-p\right) }-  \notag \\
&&\left[ \basbar{u}^{r^{\prime }}\left( q^{\prime }\right) u^{r}\left(
q\right) \right] \left[ \basbar{u}^{s^{\prime }}\left( p^{\prime }\right)
u^{s}\left( p\right) \right] e^{i\cdot x\left( q^{\prime }-q\right)
}e^{i\cdot y\left( p^{\prime }-p\right) }-  \notag \\
&&\left[ \basbar{u}^{s^{\prime }}\left( p^{\prime }\right) u^{s}\left(
p\right) \right] \left[ \basbar{u}^{r^{\prime }}\left( q^{\prime }\right)
u^{r}\left( q\right) \right] e^{i\cdot x\left( p^{\prime }-p\right)
}e^{i\cdot y\left( q^{\prime }-q\right) }+  \notag \\
&&\left[ \basbar{u}^{r^{\prime }}\left( q^{\prime }\right) u^{s}\left(
p\right) \right] \left[ \basbar{u}^{s^{\prime }}\left( p^{\prime }\right)
u^{r}\left( q\right) \right] e^{i\cdot x\left( q^{\prime }-p\right)
}e^{i\cdot y\left( p^{\prime }-q\right) }.  \label{eq7}
\end{eqnarray}%
Make use eq.(\ref{eq7}) into the $T$-matrix (\ref{eq2}) we obtain 
\begin{eqnarray*}
\left\langle f\right\vert S-1\left\vert i\right\rangle &=&\frac{\left(
-i\lambda \right) ^{2}}{2!}\int d^{4}xd^{4}y\int \frac{d^{4}k}{\left( 2\pi
\right) ^{4}}\frac{ie^{ik\cdot \left( x-y\right) }}{k^{2}-m^{2}}\times \\
&&\left\{ 
\begin{array}{c}
\left[ \basbar{u}^{s^{\prime }}\left( p^{\prime }\right) u^{r}\left(
q\right) \right] \left[ \basbar{u}^{r^{\prime }}\left( q^{\prime }\right)
u^{s}\left( p\right) \right] e^{i\cdot x\left( p^{\prime }-q\right)
}e^{i\cdot y\left( q^{\prime }-p\right) }- \\ 
\left[ \basbar{u}^{r^{\prime }}\left( q^{\prime }\right) u^{r}\left(
q\right) \right] \left[ \basbar{u}^{s^{\prime }}\left( p^{\prime }\right)
u^{s}\left( p\right) \right] e^{i\cdot x\left( q^{\prime }-q\right)
}e^{i\cdot y\left( p^{\prime }-p\right) }- \\ 
\left[ \basbar{u}^{s^{\prime }}\left( p^{\prime }\right) u^{s}\left(
p\right) \right] \left[ \basbar{u}^{r^{\prime }}\left( q^{\prime }\right)
u^{r}\left( q\right) \right] e^{i\cdot x\left( p^{\prime }-p\right)
}e^{i\cdot y\left( q^{\prime }-q\right) }+ \\ 
\left[ \basbar{u}^{r^{\prime }}\left( q^{\prime }\right) u^{s}\left(
p\right) \right] \left[ \basbar{u}^{s^{\prime }}\left( p^{\prime }\right)
u^{r}\left( q\right) \right] e^{i\cdot x\left( q^{\prime }-p\right)
}e^{i\cdot y\left( p^{\prime }-q\right) }%
\end{array}%
\right\} ,
\end{eqnarray*}%
Simplifying we obtain%
\begin{eqnarray*}
\left\langle f\right\vert S-1\left\vert i\right\rangle &=&\left( -i\lambda
\right) ^{2}(i)\int d^{4}k\left( 2\pi \right) ^{4}\times \\
&&\left\{ 
\begin{array}{c}
\frac{\left[ \basbar{u}^{r^{\prime }}\left( q^{\prime }\right) u^{s}\left(
p\right) \right] \left[ \basbar{u}^{s^{\prime }}\left( p^{\prime }\right)
u^{r}\left( q\right) \right] }{\left( p-q^{\prime }\right) ^{2}-m^{2}}\delta
^{\left( 4\right) }\left( p^{\prime }-q-p+q^{\prime }\right) - \\ 
\frac{\left[ \basbar{u}^{s^{\prime }}\left( p^{\prime }\right) u^{s}\left(
p\right) \right] \left[ \basbar{u}^{r^{\prime }}\left( q^{\prime }\right)
u^{r}\left( q\right) \right] }{\left( p-p^{\prime }\right) ^{2}-m^{2}}\delta
^{\left( 4\right) }\left( q^{\prime }-q-p+p^{\prime }\right)%
\end{array}%
\right\} , \\
&=&\left( -i\lambda \right) ^{2}(i)\int d^{4}k\left( 2\pi \right) ^{4}\delta
^{\left( 4\right) }\left( p^{\prime }+q^{\prime }-p-q\right) \times \\
&&\left\{ \frac{\left[ \basbar{u}^{r^{\prime }}\left( q^{\prime }\right)
u^{s}\left( p\right) \right] \left[ \basbar{u}^{s^{\prime }}\left(
p^{\prime }\right) u^{r}\left( q\right) \right] }{\left( p-q^{\prime
}\right) ^{2}-m^{2}}-\frac{\left[ \basbar{u}^{s^{\prime }}\left( p^{\prime
}\right) u^{s}\left( p\right) \right] \left[ \basbar{u}^{r^{\prime
}}\left( q^{\prime }\right) u^{r}\left( q\right) \right] }{\left(
p-p^{\prime }\right) ^{2}-m^{2}}\right\} ,
\end{eqnarray*}%
we find that this expression takes the form of the invariant matrix element $%
\mathcal{M}$ \cite{peskin2018}. Therefore, we have%
\begin{equation}
\mathcal{M=}\left( -i\lambda \right) ^{2}\left\{ \frac{\left[ \basbar{u}%
^{s^{\prime }}\left( p^{\prime }\right) u^{s}\left( p\right) \right] \left[ 
\basbar{u}^{r^{\prime }}\left( q^{\prime }\right) u^{r}\left( q\right) %
\right] }{\left( p-p^{\prime }\right) ^{2}-m^{2}}-\frac{\left[ \basbar{u}%
^{r^{\prime }}\left( q^{\prime }\right) u^{s}\left( p\right) \right] \left[ 
\basbar{u}^{s^{\prime }}\left( p^{\prime }\right) u^{r}\left( q\right) %
\right] }{\left( p-q^{\prime }\right) ^{2}-m^{2}}\right\} .  \label{eq8}
\end{equation}%
To compute an expression for $\left\vert \mathcal{M}\right\vert ^{2}$ we
need to multiply $\mathcal{M}$ by the complex conjugate $\mathcal{M}^{\ast }$%
\begin{eqnarray*}
\mathcal{MM}^{\ast } &=&\left( -i\lambda \right) ^{2}\left\{ \frac{\left[ 
\basbar{u}^{s^{\prime }}\left( p^{\prime }\right) u^{s}\left( p\right) %
\right] \left[ \basbar{u}^{r^{\prime }}\left( q^{\prime }\right)
u^{r}\left( q\right) \right] }{\left( p-p^{\prime }\right) ^{2}-m^{2}}-\frac{%
\left[ \basbar{u}^{r^{\prime }}\left( q^{\prime }\right) u^{s}\left(
p\right) \right] \left[ \basbar{u}^{s^{\prime }}\left( p^{\prime }\right)
u^{r}\left( q\right) \right] }{\left( p-q^{\prime }\right) ^{2}-m^{2}}%
\right\} \times \\
&&\left( i\lambda \right) ^{2}\left\{ \frac{\left[ \basbar{u}^{s^{\prime
}}\left( p^{\prime }\right) u^{s}\left( p\right) \right] ^{\ast }\left[ 
\basbar{u}^{r^{\prime }}\left( q^{\prime }\right) u^{r}\left( q\right) %
\right] ^{\ast }}{\left( p-p^{\prime }\right) ^{2}-m^{2}}-\frac{\left[ 
\basbar{u}^{r^{\prime }}\left( q^{\prime }\right) u^{s}\left( p\right) %
\right] ^{\ast }\left[ \basbar{u}^{s^{\prime }}\left( p^{\prime }\right)
u^{r}\left( q\right) \right] ^{\ast }}{\left( p-q^{\prime }\right) ^{2}-m^{2}%
}\right\} .
\end{eqnarray*}%
Since the Nucleon beams are unpolarized the measured Cross-section is an
average over the initial spins $s,r$ \ and a sum over the final spins $%
s^{\prime },r^{\prime }$, we could compute $\left\vert \mathcal{M}\left(
s,r\longrightarrow s^{\prime },r^{\prime }\right) \right\vert ^{2}$ such as 
\begin{eqnarray}
\left\langle \left\vert \mathcal{M}\right\vert ^{2}\right\rangle &=&\frac{1}{%
2}\sum_{s}\frac{1}{2}\sum_{r}\sum_{s^{\prime }}\sum_{r^{\prime }}\left\vert 
\mathcal{M}\left( s,r\longrightarrow s^{\prime },r^{\prime }\right)
\right\vert ^{2}  \notag \\
&=&\frac{\lambda ^{4}}{4}\sum_{s,r,s^{\prime },r^{\prime }}\left\{ 
\begin{array}{c}
\frac{\left[ \basbar{u}^{s^{\prime }}\left( p^{\prime }\right) u^{s}\left(
p\right) \right] \left[ \basbar{u}^{r^{\prime }}\left( q^{\prime }\right)
u^{r}\left( q\right) \right] \left[ \basbar{u}^{s^{\prime }}\left(
p^{\prime }\right) u^{s}\left( p\right) \right] ^{\ast }\left[ \basbar{u}%
^{r^{\prime }}\left( q^{\prime }\right) u^{r}\left( q\right) \right] ^{\ast }%
}{\left[ \left( p-p^{\prime }\right) ^{2}-m^{2}\right] ^{2}}- \\ 
\frac{\left[ \basbar{u}^{s^{\prime }}\left( p^{\prime }\right) u^{s}\left(
p\right) \right] \left[ \basbar{u}^{r^{\prime }}\left( q^{\prime }\right)
u^{r}\left( q\right) \right] \left[ \basbar{u}^{r^{\prime }}\left(
q^{\prime }\right) u^{s}\left( p\right) \right] ^{\ast }\left[ \basbar{u}%
^{s^{\prime }}\left( p^{\prime }\right) u^{r}\left( q\right) \right] ^{\ast }%
}{\left[ \left( p-p^{\prime }\right) ^{2}-m^{2}\right] \left[ \left(
p-q^{\prime }\right) ^{2}-m^{2}\right] }- \\ 
\frac{\left[ \basbar{u}^{r^{\prime }}\left( q^{\prime }\right) u^{s}\left(
p\right) \right] \left[ \basbar{u}^{s^{\prime }}\left( p^{\prime }\right)
u^{r}\left( q\right) \right] \left[ \basbar{u}^{s^{\prime }}\left(
p^{\prime }\right) u^{s}\left( p\right) \right] ^{\ast }\left[ \basbar{u}%
^{r^{\prime }}\left( q^{\prime }\right) u^{r}\left( q\right) \right] ^{\ast }%
}{\left[ \left( p-q^{\prime }\right) ^{2}-m^{2}\right] \left[ \left(
p-p^{\prime }\right) ^{2}-m^{2}\right] }+ \\ 
\frac{\left[ \basbar{u}^{r^{\prime }}\left( q^{\prime }\right) u^{s}\left(
p\right) \right] \left[ \basbar{u}^{s^{\prime }}\left( p^{\prime }\right)
u^{r}\left( q\right) \right] \left[ \basbar{u}^{r^{\prime }}\left(
q^{\prime }\right) u^{s}\left( p\right) \right] ^{\ast }\left[ \basbar{u}%
^{s^{\prime }}\left( p^{\prime }\right) u^{r}\left( q\right) \right] ^{\ast }%
}{\left[ \left( p-q^{\prime }\right) ^{2}-m^{2}\right] ^{2}}%
\end{array}%
\right\} .  \label{eq9}
\end{eqnarray}%
we can perform the spin sum using the completeness relation \cite{peskin2018}
\begin{equation*}
\sum_{s}u^{s}\left( p\right) \basbar{u}^{s}\left( p\right) =\left( \gamma
^{\mu }p_{\mu }+M\right) .
\end{equation*}%
where $s=1,2$ for the two spin state, and $\gamma ^{\mu }p_{\mu }=\slashed{p}
$ , and the adjoint, $\basbar{u}=u^{\dagger }\gamma ^{0}$. Using the
Casimir's trick reduces everything to a trace for the product of gamma
matrices. Evaluating these traces using the algebraic properties for the
of the gamma matrices trace rules \cite{griffiths2020}, we arrive to the scattering
amplitude
\begin{equation}
\left\langle \left\vert \mathcal{M}\right\vert ^{2}\right\rangle =\frac{%
\lambda ^{4}}{4}\left\{ 
\begin{array}{c}
16 \frac{\left( p\cdot p^{\prime }+M^{2}\right) \left( q^{\prime
}\cdot q+M^{2}\right) }{\left[ \left( p-p^{\prime }\right) ^{2}-m^{2}\right]
^{2}} - \\ 
8 \frac{\left( p^{\prime }\cdot p\right) \left( q^{\prime }\cdot
q\right) -\left( p^{\prime }\cdot q^{\prime }\right) \left( p\cdot q\right)
+\left( p^{\prime }\cdot q\right) \left( p\cdot q^{\prime }\right)
+M^{2}\left( p^{\prime }\cdot q+p\cdot q^{\prime }+p^{\prime }\cdot
q^{\prime }+p^{\prime }\cdot p+q^{\prime }\cdot q+p\cdot q\right) +M^{4}}{%
\left[ \left( p-p^{\prime }\right) ^{2}-m^{2}\right] \left[ \left(
p-q^{\prime }\right) ^{2}-m^{2}\right] }+ \\ 
+16 \frac{\left( p\cdot q^{\prime }+M^{2}\right) \left( p^{\prime
}\cdot q+M^{2}\right) }{\left[ \left( p-q^{\prime }\right) ^{2}-m^{2}\right]
^{2}}%
\end{array}%
\right\} .  \label{eq10}
\end{equation}

\subsection{Elastic scattering}

For a more explicit formula, we express the 4-momentum vectors in terms of
energies and angles in the center-of-mass frame to evaluate the
Cross-section. This can be drawn as follows

\tikzset{every picture/.style={line width=0.75pt}} 
\begin{tikzpicture}[x=0.75pt,y=0.75pt,yscale=-1,xscale=1]
\draw    (343.62,129.21) -- (411.41,130.01) ;
\draw [shift={(340.62,129.18)}, rotate = 0.67] [fill={rgb, 255:red, 0; green, 0; blue, 0 }  ][line width=0.08]  [draw opacity=0] (8.93,-4.29) -- (0,0) -- (8.93,4.29) -- cycle    ;
\draw    (306.6,129.18) -- (238.81,129.18) ;
\draw [shift={(309.6,129.18)}, rotate = 180] [fill={rgb, 255:red, 0; green, 0; blue, 0 }  ][line width=0.08]  [draw opacity=0] (8.93,-4.29) -- (0,0) -- (8.93,4.29) -- cycle    ;
\draw    (367.92,63.18) -- (331.25,119.52) ;
\draw [shift={(369.56,60.67)}, rotate = 123.06] [fill={rgb, 255:red, 0; green, 0; blue, 0 }  ][line width=0.08]  [draw opacity=0] (8.93,-4.29) -- (0,0) -- (8.93,4.29) -- cycle    ;
\draw    (279.87,198.83) -- (318.58,140.19) ;
\draw [shift={(278.22,201.33)}, rotate = 303.43] [fill={rgb, 255:red, 0; green, 0; blue, 0 }  ][line width=0.08]  [draw opacity=0] (8.93,-4.29) -- (0,0) -- (8.93,4.29) -- cycle    ;
\draw  [draw opacity=0] (340.35,105.16) .. controls (341.34,103.86) and (342.62,102.86) .. (344.16,102.29) .. controls (349.64,100.23) and (356.29,104.41) .. (359,111.62) .. controls (361.71,118.83) and (359.46,126.33) .. (353.97,128.38) .. controls (352.43,128.96) and (350.81,129.04) .. (349.2,128.71) -- (349.06,115.33) -- cycle ; \draw   (340.35,105.16) .. controls (341.34,103.86) and (342.62,102.86) .. (344.16,102.29) .. controls (349.64,100.23) and (356.29,104.41) .. (359,111.62) .. controls (361.71,118.83) and (359.46,126.33) .. (353.97,128.38) .. controls (352.43,128.96) and (350.81,129.04) .. (349.2,128.71) ;  
\draw (368.08,104.11) node [anchor=north west][inner sep=0.75pt]    {$\theta _{cm}$};
\draw (223.56,123.08) node [anchor=north west][inner sep=0.75pt]    {$p$};
\draw (418.9,123.24) node [anchor=north west][inner sep=0.75pt]    {$q$};
\draw (374.58,41.91) node [anchor=north west][inner sep=0.75pt]    {$p^{\prime }$};
\draw (272.04,206.91) node [anchor=north west][inner sep=0.75pt]    {$q^{\prime }$};
\end{tikzpicture}

The center-of-mass frame sets that 
\begin{equation*}
\mathbf{p=-q}\text{\quad and\quad }\mathbf{p}^{\prime }\mathbf{=-q}^{\prime }
\end{equation*}%
where%
\begin{eqnarray*}
p &=&\left( E,\mathbf{p}\right) ;\qquad q=\left( E,-\mathbf{p}\right) , \\
p^{\prime } &=&\left( E,\mathbf{p}^{\prime }\right) ;\qquad q^{\prime
}=\left( E,-\mathbf{p}^{\prime }\right) ,
\end{eqnarray*}%
where $E$ \ is the single nucleon energy in the center of the mass frame. Thus%
\begin{equation}
E_{cm}=(p+q)=2E=\sqrt{2M^{2}+2ME_{p}^{lab}}=\sqrt{s}.  \label{eq11}
\end{equation}%
where $E_{cm}$ is the total center of mass energy of the scattering.
Recall that%
\begin{equation}
E_{p}^{lab}=T_{lab}+M,  \label{eq11_1}
\end{equation}%
where $M\approx 939$ MeV and $E_{lap}<2m_{\pi }c^{2}$ whereas $m_{\pi
}=134.977$ MeV, Since we need the lab energy below the threshold energy of
pion formation. We conclude that for the elastic scattering in the center of
mass, the momentum satisfies%
\begin{equation*}
\left\vert \mathbf{p}\right\vert =\left\vert \mathbf{p}^{\prime }\right\vert
,
\end{equation*}%
meaning the change in momentum is just in direction, not in magnitude. Thus,%
\begin{equation}
\mathbf{p\cdot p}^{\prime }=\left\vert \mathbf{p}\right\vert ^{2}\cos \theta
\label{eq11_2}
\end{equation}%
where $\theta $ is the center of mass scattering angle and%
\begin{equation*}
\left\vert \mathbf{p}\right\vert ^{2}=E^{2}-M^{2}
\end{equation*}%
For NN scattering we have $M_{p}=M_{q}=M_{p^{\prime }}=M_{q^{\prime
}}=M\approx 939$ MeV. Therefore We can rewrite eq.(\ref{eq10}) in terms of $%
E $ and $\theta $ :%
\begin{equation}
\left\langle \left\vert \mathcal{M}\right\vert ^{2}\right\rangle =\frac{%
\lambda ^{4}}{4}\left\{ 
\begin{array}{c}
\frac{16\left( \left( E^{2}-M^{2}\right) \cos \theta +E^{2}+M^{2}\right) ^{2}%
}{\left( 4\left( E^{2}-M^{2}\right) \cos ^{2}\frac{\theta }{2}+m^{2}\right)
^{2}}+ \\ 
\frac{16\left( \left( M^{2}-E^{2}\right) \cos \theta +E^{2}+M^{2}\right) ^{2}%
}{\left( 4\left( E^{2}-M^{2}\right) \sin ^{2}\frac{\theta }{2}+m^{2}\right)
^{2}}- \\ 
\frac{8\left( -E^{4}+\left( E^{2}-M^{2}\right) ^{2}\cos (2\theta
)+10E^{2}M^{2}-M^{4}\right) }{\left( 2E^{2}+m^{2}-2M^{2}\right) ^{2}-4\left(
E^{2}-M^{2}\right) ^{2}\cos ^{2}\theta }%
\end{array}%
\right\} .  \label{eq12}
\end{equation}%
where $\lambda ^{4}$ is the nuclear coupling constant. also, eq.(\ref{eq10})
can be written in terms of Mandelstam variables instead of the energy and
angle to have a simpler form for the differential cross-section \cite{rpp_pdg_2022}%
\begin{eqnarray}
s &=&\left( p+q\right) ^{2}=\left( p^{\prime }+q^{\prime }\right)
^{2}=4E^{2};  \notag \\
t &=&\left( p-p^{\prime }\right) ^{2}=\left( q^{\prime }-q\right)
^{2}=-4\left( E^{2}-M^{2}\right) \sin \left( \frac{\theta }{2}\right) ^{2}; 
\notag \\
u &=&\left( p-q^{\prime }\right) ^{2}=\left( p^{\prime }-q\right)
^{2}=-4\left( E^{2}-M^{2}\right) \cos \left( \frac{\theta }{2}\right) ^{2},
\label{eq13}
\end{eqnarray}%
The Mandelstam variables $s$, $t$, and $u$ are invariant, so if we express
the amplitude and cross-section in terms of Mandelstam variables we should
have the same answer regardless of what kind of metric we use. The sum of
the Mandelstam variables is%
\begin{equation}
s+t+u=4M^{2}, \label{eq13_1}
\end{equation}%
when all four particles have a mass of $M$. Thus the amplitude in terms of
Mandelstam variables becomes%
\begin{equation}
\left\langle \left\vert \mathcal{M}\right\vert ^{2}\right\rangle =\lambda
^{4}\left[ \frac{\left( s+u\right) ^{2}}{\left( t-m^{2}\right)^{2}}-\frac{%
s^{2}+s\left( t+u\right) }{\left(t-m^{2}\right) \left(u-m^{2}\right) }+%
\frac{\left( s+t\right) ^{2}}{\left( u-m^{2}\right) ^{2}}\right] .
\label{eq13_3}
\end{equation}%

\subsection{Nucleon-nucleon scattering using pseudoscalar current}

Yukawa originally proposed an interaction as an effective theory of nuclear
forces. With an eye to this, we will again refer to the $\phi $ particles as
scalar mesons (pions), and the $\psi $ particles as nucleons. The nucleons
have spin and isospin. Moreover, in Nature the relevant mesons are pions
which are pseudoscalar, so a coupling of the form $\phi \basbar{\psi }%
\gamma ^{5}\psi $ would seem to be more appropriate. Thus, we re-write our
Lagrangian eq.(\ref{eq1}) to be%
\begin{equation}
\mathcal{L=}\frac{1}{2}\partial _{\mu }\phi \partial ^{\mu }\phi -\frac{1}{2}%
m^{2}\phi ^{2}+\basbar{\psi }\left( i\gamma ^{\mu }\partial _{\mu
}-M\right) \psi -\lambda \phi \basbar{\psi }\gamma ^{5}\psi .  \label{eq15}
\end{equation}%
The interaction vertex $-i\lambda $ is now changed to a factor of $-i\lambda
\gamma ^{5}$using the Feynman rule \cite{griffiths2020,rpp_pdg_2022}. Now,
the amplitude with a pseudoscalar coupling is%
\begin{equation}
\mathcal{M=}\left( -i\lambda \right) ^{2}\left\{ \frac{\left[ \basbar{u}%
^{s^{\prime }}\left( p^{\prime }\right) \gamma ^{5}u^{s}\left( p\right) %
\right] \left[ \basbar{u}^{r^{\prime }}\left( q^{\prime }\right) \gamma
^{5}u^{r}\left( q\right) \right] }{\left( p-p^{\prime }\right) ^{2}-m^{2}}-%
\frac{\left[ \basbar{u}^{r^{\prime }}\left( q^{\prime }\right) \gamma
^{5}u^{s}\left( p\right) \right] \left[ \basbar{u}^{s^{\prime }}\left(
p^{\prime }\right) \gamma ^{5}u^{r}\left( q\right) \right] }{\left(
p-q^{\prime }\right) ^{2}-m^{2}}\right\} .  \label{eq16}
\end{equation}%
The only modification is that we added a factor of $\gamma ^{5}$. Multiply $%
\mathcal{M}$ by $\mathcal{M}^{\ast }$, we have the squared scattering
amplitude%
\begin{equation}
\left\langle \left\vert \mathcal{M}\right\vert ^{2}\right\rangle =\frac{%
\lambda ^{4}}{4}\sum_{s,r,s^{\prime },r^{\prime }}\left\{ 
\begin{array}{c}
\frac{\left[ \basbar{u}^{s^{\prime }}\left( p^{\prime }\right) \gamma
^{5}u^{s}\left( p\right) \right] \left[ \basbar{u}^{r^{\prime }}\left(
q^{\prime }\right) \gamma ^{5}u^{r}\left( q\right) \right] \left[ \basbar{u%
}^{s^{\prime }}\left( p^{\prime }\right) \gamma ^{5}u^{s}\left( p\right) %
\right] ^{\ast }\left[ \basbar{u}^{r^{\prime }}\left( q^{\prime }\right)
\gamma ^{5}u^{r}\left( q\right) \right] ^{\ast }}{\left[ \left( p-p^{\prime
}\right) ^{2}-m^{2}\right] ^{2}}- \\ 
\frac{\left[ \basbar{u}^{s^{\prime }}\left( p^{\prime }\right) \gamma
^{5}u^{s}\left( p\right) \right] \left[ \basbar{u}^{r^{\prime }}\left(
q^{\prime }\right) \gamma ^{5}u^{r}\left( q\right) \right] \left[ \basbar{u%
}^{r^{\prime }}\left( q^{\prime }\right) \gamma ^{5}u^{s}\left( p\right) %
\right] ^{\ast }\left[ \basbar{u}^{s^{\prime }}\left( p^{\prime }\right)
\gamma ^{5}u^{r}\left( q\right) \right] ^{\ast }}{\left[ \left( p-p^{\prime
}\right) ^{2}-m^{2}\right] \left[ \left( p-q^{\prime }\right) ^{2}-m^{2}%
\right] }- \\ 
\frac{\left[ \basbar{u}^{r^{\prime }}\left( q^{\prime }\right) \gamma
^{5}u^{s}\left( p\right) \right] \left[ \basbar{u}^{s^{\prime }}\left(
p^{\prime }\right) \gamma ^{5}u^{r}\left( q\right) \right] \left[ \basbar{u%
}^{s^{\prime }}\left( p^{\prime }\right) \gamma ^{5}u^{s}\left( p\right) %
\right] ^{\ast }\left[ \basbar{u}^{r^{\prime }}\left( q^{\prime }\right)
\gamma ^{5}u^{r}\left( q\right) \right] ^{\ast }}{\left[ \left( p-q^{\prime
}\right) ^{2}-m^{2}\right] \left[ \left( p-p^{\prime }\right) ^{2}-m^{2}%
\right] }+ \\ 
\frac{\left[ \basbar{u}^{r^{\prime }}\left( q^{\prime }\right) \gamma
^{5}u^{s}\left( p\right) \right] \left[ \basbar{u}^{s^{\prime }}\left(
p^{\prime }\right) \gamma ^{5}u^{r}\left( q\right) \right] \left[ \basbar{u%
}^{r^{\prime }}\left( q^{\prime }\right) \gamma ^{5}u^{s}\left( p\right) %
\right] ^{\ast }\left[ \basbar{u}^{s^{\prime }}\left( p^{\prime }\right)
\gamma ^{5}u^{r}\left( q\right) \right] ^{\ast }}{\left[ \left( p-q^{\prime
}\right) ^{2}-m^{2}\right] ^{2}}%
\end{array}%
\right\} .  \label{eq17}
\end{equation}%
where $\basbar{\gamma ^{5}}=\gamma ^{0}\gamma ^{5}\gamma ^{0}=-\gamma ^{5}$%
, $\left( \gamma ^{5}\right) ^{2}=1$ and $\gamma ^{5}$ anti-commutes with $%
\gamma $ matrices $\left\{ \gamma ^{\mu },\gamma ^{5}\right\} =0$ \cite%
{peskin2018,rpp_pdg_2022}. It is much easier to compute the squared
scattering amplitude directly using general Casimir's Trick. By applying
this trick and using the algebraic properties for the trace of gamma
matrices we get the final amplitude%
\begin{equation}
\left\langle \left\vert \mathcal{M}\right\vert ^{2}\right\rangle =\frac{%
\lambda ^{4}}{4}\left\{ 
\begin{array}{c}
16\left\{ \frac{\left( p\cdot p^{\prime }-M^{2}\right) \left( q\cdot
q^{\prime }-M^{2}\right) }{\left[ \left( p-p^{\prime }\right) ^{2}-m^{2}%
\right] ^{2}}\right\} - \\ 
8\left\{ \frac{\left[ \left( p^{\prime }\cdot p\right) \left( q^{\prime
}\cdot q\right) -\left( p^{\prime }\cdot q^{\prime }\right) \left( p\cdot
q\right) +\left( p^{\prime }\cdot q\right) \left( p\cdot q^{\prime }\right) %
\right] +M^{2}\left( p^{\prime }\cdot q^{\prime }+p\cdot q-q^{\prime }\cdot
q-p\cdot q^{\prime }-p^{\prime }\cdot p-p^{\prime }\cdot q\right) +M^{4}}{%
\left[ \left( p-p^{\prime }\right) ^{2}-m^{2}\right] \left[ \left(
p-q^{\prime }\right) ^{2}-m^{2}\right] }\right\} + \\ 
16\left\{ \frac{\left( p\cdot q^{\prime }-M^{2}\right) \left( q\cdot
p^{\prime }-M^{2}\right) }{\left[ \left( p-q^{\prime }\right) ^{2}-m^{2}%
\right] ^{2}}\right\}%
\end{array}%
\right\} .  \label{eq18}
\end{equation}%
In terms of $E$ and $\theta $ 
\begin{equation}
\left\langle \left\vert \mathcal{M}\right\vert ^{2}\right\rangle =\frac{%
\lambda ^{4}}{4}\frac{%
\begin{array}{c}
8\left( E^{2}-M^{2}\right) ^{2}\times  \\ 
\left( 
\begin{array}{c}
9E^{4}+12E^{2}m^{2}+3\left( E^{2}-M^{2}\right) ^{2}\cos (4\theta
)-18E^{2}M^{2}+ \\ 
\cos (2\theta )\left( -12E^{4}-12E^{2}\left( m^{2}-2M^{2}\right)
+m^{4}+12m^{2}M^{2}-12M^{4}\right) + \\ 
7m^{4}-12m^{2}M^{2}+9M^{4}%
\end{array}%
\right) 
\end{array}%
}{\left[ \left( 2E^{2}+m^{2}-2M^{2}\right) ^{2}-4\left( E^{2}-M^{2}\right)
^{2}\cos ^{2}(\theta )\right] ^{2}}.  \label{eq49_1}
\end{equation}%
In terms of Mandelstam variable becomes%
\begin{equation}
\left\langle \left\vert \mathcal{M}\right\vert ^{2}\right\rangle =\lambda
^{4}\left[ \frac{t^{2}}{\left( t-m^{2}\right) ^{2}}+\frac{tu}{\left(
t-m^{2}\right) \left( u-m^{2}\right) }+\frac{u^{2}}{\left( u-m^{2}\right)
^{2}}\right]  \label{eq.ampl_mandilstan}
\end{equation}%

\subsection{Inelastic scattering}

In general case scattering%
\begin{equation}
1+2\rightarrow 3+4, \label{eqscattering}
\end{equation}%
we have the following 4 component momenta in the center of mass frame%
\begin{equation}
p_{1}=(E_{1},\mathbf{p})=(E_{1},\sqrt{E_{1}^{2}-M_{1}^{2}})\hat{\mathbf{z}},
\label{eqp1}
\end{equation}%
\begin{eqnarray}
p_{2} &=&(E_{2},-\mathbf{p})=(E_{2},-\sqrt{E_{2}^{2}-M_{2}^{2}})\hat{\mathbf{%
z}},  \notag \\
&=&(E_{2},-\sqrt{E_{1}^{2}-M_{1}^{2}})\hat{\mathbf{z}},  \label{eqp2}
\end{eqnarray}%
\begin{eqnarray}
p_{3} &=&(E_{3},\mathbf{p}^{\prime });  \notag \\
&=&(E_{3},\sqrt{E_{3}^{2}-M_{3}^{2}}\sin \theta \cos \phi ,\sqrt{%
E_{3}^{2}-M_{3}^{2}}\sin \theta \sin \phi ,\sqrt{E_{3}^{2}-M_{3}^{2}}\cos
\theta ),  \label{eqp3}
\end{eqnarray}%
and%
\begin{eqnarray}
p_{4} &=&(E_{4},-\mathbf{p}^{\prime });  \notag \\
&=&(E_{4},-\sqrt{E_{4}^{2}-M_{4}^{2}}\sin \theta \cos \phi ,-\sqrt{%
E_{4}^{2}-M_{4}^{2}}\sin \theta \sin \phi ,-\sqrt{E_{4}^{2}-M_{4}^{2}}\cos
\theta ),  \notag \\
&=&(E_{4},-\sqrt{E_{3}^{2}-M_{3}^{2}}\sin \theta \cos \phi ,-\sqrt{%
E_{3}^{2}-M_{3}^{2}}\sin \theta \sin \phi ,-\sqrt{E_{3}^{2}-M_{3}^{2}}\cos
\theta ).  \label{eqp4}
\end{eqnarray}%
Conservation of linear momentum requires%
\begin{equation}
\mathbf{p}_{1}=-\mathbf{p}_{2}\quad \Rightarrow
E_{1}^{2}-M_{1}^{2}=E_{2}^{2}-M_{2}^{2}.  \label{eqconsmom1}
\end{equation}%
Special case, $M_{1}=M_{2}$ we reach to%
\begin{equation}
E_{1}=E_{2}.  \label{eqequalmass1}
\end{equation}%
similarly%
\begin{equation}
E_{3}^{2}-M_{3}^{2}=E_{4}^{2}-M_{4}^{2}.  \label{eqconsmom2}
\end{equation}%
The quantity%
\begin{equation}
p_{1}+p_{2}=(E_{1}+E_{2},\mathbf{0}),  \label{eqp1plusp2}
\end{equation}%
is not Lorentz invariant. The contraction, however, is given by%
\begin{eqnarray}
s &\equiv &(p_{1}+p_{2})^{2}=(p_{3}+p_{4})^{2}  \label{eqmandelsans} \\
&=&(E_{1}+E_{2})^{2}=(E_{3}+E_{4})^{2}.
\end{eqnarray}%
Whereas the $t$ and $u$ channels are%
\begin{eqnarray*}
t &=&\left( p_{1}-p_{3}\right) ^{2}=\left( p_{2}-p_{4}\right) ^{2}; \\
&=&(E_{1}-E_{3})^{2}-(\mathbf{p}-\mathbf{p}^{\prime
})^{2}=(E_{2}-E_{4})^{2}-(\mathbf{p}^{\prime }-\mathbf{p})^{2}.
\end{eqnarray*}%
and%
\begin{eqnarray*}
u &=&\left( p_{1}-p_{4}\right) ^{2}=\left( p_{2}-p_{3}\right) ^{2}; \\
&=&(E_{1}-E_{4})^{2}-(\mathbf{p}+\mathbf{p}^{\prime
})^{2}=(E_{2}-E_{3})^{2}-(\mathbf{p}+\mathbf{p}^{\prime })^{2}
\end{eqnarray*}%
We can thus express $E_{1}$ and $E_{3}$ in terms of the invariants as \cite%
{HS,rpp_pdg_2022}%
\begin{equation}
E_{1,3}=\frac{1}{2\sqrt{s}}\left( s+M_{1,3}^{2}-M_{2,4}^{2}\right) .
\label{eqenergy13}
\end{equation}%
\begin{eqnarray}
E_{3} &=&\frac{1}{2(E_{1}+E_{2})}\left(
(E_{1}+E_{2})^{2}+M_{3}^{2}-M_{4}^{2}\right)  \notag \\
&=&\frac{1}{4E}\left( 4E^{2}+M_{3}^{2}-M_{4}^{2}\right) .  \label{eqe3}
\end{eqnarray}%
We can avoid using $E_{4}$ in the invariant scattering amplitudes eqs.
(\ref{eq13_3}) and (\ref{eq.ampl_mandilstan}), by using%
\begin{equation*}
t=\left( p_{1}-p_{3}\right) ^{2};\quad \mathrm{and}\quad u=\left(
p_{2}-p_{3}\right) ^{2}.
\end{equation*}%
The special case when the incident particle in the lab frame has energy $%
E_{lab}=T_{lab}+M$ incident on an equal mass target. In this case%
\begin{equation*}
M_{1}=M_{2}=M,\qquad \text{and\qquad }E_{1}=E_{2}=E.
\end{equation*}%
Here $E_{cm}=2E$. In inelastic scattering, the center of mass energy must exceed
the final mass, i.e.%
\begin{equation}
E_{cm}=\sqrt{M_{1}^{2}+M_{1}^{2}+2M_{2}E_{lab}}>M_{3}+M_{4}.
\label{eqecmthreshold}
\end{equation}%
In our special case%
\begin{equation*}
\sqrt{2M^{2}+2M\left( T_{lab}+M\right) }>M_{3}+M_{4},
\end{equation*}%
to reach threshold for the $T_{lab}$ to be%
\begin{equation}
T_{lab}>\frac{(M_{3}+M_{4})^{2}-4M^{2}}{2M}.  \label{eqtlabthreshold}
\end{equation}

\subsection{Cross section}

The final form of the relation between $S$-Matrix elements and cross-section 
\cite{peskin2018} 
\begin{eqnarray}
d\sigma  &=&\frac{1}{4E_{T}E_{I}\left\vert v_{T}-v_{I}\right\vert }\left(
\prod\limits_{f}\frac{d^{3}p_{f}}{\left( 2\pi \right) ^{3}}\frac{1}{2E_{f}}%
\right) \times   \notag \\
&&\left\vert \mathcal{M}\left( p_{T},p_{I}\longrightarrow \left\{
p_{f}\right\} \right) \right\vert ^{2}\left( 2\pi \right) ^{4}\delta
^{\left( 4\right) }\left( P_{T}+P_{I}-\sum p_{f}\right) ,  \label{eqCS}
\end{eqnarray}%
where the indices $I$ and $T$ denotes the incident and target particle. The
shape of the wave packets is practically irrelevant. We can simplify this
expression by the integral over two-particle in the final state in the
center-of-mass frame which is sets that $p_{2}=-p_{1}$ , The integral over
two-body phase space becomes%
\begin{equation}
\int d\Pi _{2}=\int dp_{1}p_{1}^{2}d\Omega \frac{dp_{1}p_{1}^{2}d\Omega }{%
\left( 2\pi \right) ^{3}2E_{1}2E_{2}}\left( 2\pi \right) \delta \left(
E_{cm}-E_{1}-E_{2}\right)   \label{eqFinalIntegral}
\end{equation}%
Where $E_{cm}$ is the total initial energy and $E_{1}=\sqrt{%
p_{1}^{2}+m_{1}^{2}}$ , $E_{2}=\sqrt{p_{2}^{2}+m_{2}^{2}}$ , taking the
integral over the final delta function 
\begin{eqnarray}
\int d\Pi _{2} &=&\int d\Omega \frac{p_{1}^{2}}{16\pi ^{2}E_{1}E_{2}}\left( 
\frac{p_{1}}{E_{1}}+\frac{p_{1}}{E_{2}}\right) ^{-1}  \notag \\
&=&\int d\Omega \frac{1}{16\pi ^{2}}\frac{\left\vert p_{1}\right\vert }{%
E_{cm}}.
\end{eqnarray}%
Applying these simplifications to eq. (\ref{eqCS}), we get the cross-section for
two final-state particles%
\begin{equation}
\left( \frac{d\sigma }{d\Omega }\right) _{cm}=\frac{1}{4E_{T}E_{I}\left\vert
v_{T}-v_{I}\right\vert }\frac{\left\vert \mathbf{p}_{I}\right\vert }{\left(
2\pi \right) ^{2}4E_{cm}}\left\vert \mathcal{M}\left(
p_{T},p_{I}\longrightarrow p_{1},p_{2}\right) \right\vert ^{2}.  \label{eq19}
\end{equation}%
The relative velocity factor can be transformed as%
\begin{equation}
\left\vert v_{T}-v_{I}\right\vert =\left\vert \frac{\mathbf{p}_{T}}{E_{T}}-%
\frac{\mathbf{p}_{I}}{E_{I}}\right\vert =\frac{\left\vert \mathbf{p}%
_{I}\right\vert }{E_{T}E_{I}}\underset{E_{cm}=\sqrt{s}}{\underbrace{\left(
E_{T}+E_{I}\right) }}.  \label{eqvrel}
\end{equation}%
Make use eq. (\ref{eqvrel}) into (\ref{eq19}), we obtain the cross-section
formula in center of mass frame,%
\begin{equation}
\left( \frac{d\sigma }{d\Omega }\right) _{cm}=\frac{\left\vert \mathcal{M}%
\right\vert ^{2}}{64\pi ^{2}E_{cm}^{2}}.  \label{eq20}
\end{equation}%
Now, we can plug the amplitude (\ref{eq13_3}) into the differential
cross-section (\ref{eq20}) to obtain the differential cross-section for NN
scattering using scalar current 
\begin{equation}
\left( \frac{d\sigma }{d\Omega }\right) _{CM}=\frac{\lambda ^{4}}{64\pi ^{2}s%
}\left[ \frac{\left( s+u\right) ^{2}}{\left( t-m^{2}\right) ^{2}}-\frac{%
s^{2}+s\left( t+u\right) }{\left( t-m^{2}\right) \left( u-m^{2}\right) }+%
\frac{\left( s+t\right) ^{2}}{\left( u-m^{2}\right) ^{2}}\right] .
\label{eq14}
\end{equation}
Similarly, plugging the amplitude (\ref{eq.ampl_mandilstan}) into the differential
cross-section (\ref{eq20}) we obtain the differential cross-section for NN
scattering using pseudoscalar current becomes%
\begin{equation}
\left( \frac{d\sigma }{d\Omega }\right) _{cm}=\frac{\lambda ^{4}}{64\pi
^{2}s}\left[ \frac{t^{2}}{\left( t-m^{2}\right) ^{2}}+\frac{tu}{\left(
t-m^{2}\right) \left( u-m^{2}\right) }+\frac{u^{2}}{\left( u-m^{2}\right)
^{2}}\right] .  \label{eqDCSgamma5}
\end{equation}%
The total cross-section is obtained by integrating the differential
cross-section over $d\Omega $%
\begin{equation}
\sigma =\int \frac{d\sigma }{d\Omega }d\Omega   \label{eqCS1}
\end{equation}%
where $d\Omega =\sin \theta d\theta d\phi $ is the solid angle, thus the
total cross-section is%
\begin{equation*}
\sigma =\int \frac{d\sigma }{d\Omega }\sin \theta d\theta d\phi ,
\end{equation*}%
In our case, there is no azimuthal dependent on the differential cross
section. Thus the total cross-section becomes%
\begin{equation}
\sigma =2\pi \int\limits_{0}^{\pi }\frac{d\sigma }{d\Omega }\sin \theta
d\theta .  \label{eqCS3}
\end{equation}

\subsection{Differential cross section in laboratory frame}

To compare the theoretical predictions with the experimental values, we need
to convert the center of the mass differential cross-section into the laboratory
frame one. This is achieved by the fact that%
\begin{equation}
d\sigma _{lab}=d\sigma _{cm}\quad \Rightarrow \quad \left( \frac{d\sigma }{%
d\Omega }\right) _{lab}d\Omega _{lab}=\left( \frac{d\sigma }{d\Omega }%
\right) _{cm}d\Omega _{cm}.  \label{eqDCSlabtoCM}
\end{equation}%
Therefore,%
\begin{eqnarray*}
\left( \frac{d\sigma }{d\Omega }\right) _{lab} &=&\left( \frac{d\sigma }{%
d\Omega }\right) _{cm}\frac{d\Omega _{cm}}{d\Omega _{lab}} \\
&=&\left( \frac{d\sigma }{d\Omega }\right) _{cm}\frac{\sin \theta
_{cm}d\theta _{cm}d\phi _{cm}}{\sin \theta _{lab}d\theta _{lab}d\phi _{lab}}.
\end{eqnarray*}%
In the case of azimuthal symmetry, angels $\phi _{cm}=\phi _{lab}$. The
differential cross section in the lab frame becomes%
\begin{equation}
\left( \frac{d\sigma }{d\Omega }\right) _{lab}=\left( \frac{d\sigma }{%
d\Omega }\right) _{cm}\frac{d(\cos \theta _{cm})}{d(\cos \theta _{lab})},
\label{eqDCSlabtoCM2}
\end{equation}%
where the differential factor $d(\cos \theta _{cm})/d(\cos \theta _{lab})$
is calculated from \cite{HS}%
\begin{equation}
\tan \theta _{lab}=\frac{\sin \theta _{cm}}{\gamma (\cos \theta
_{cm}+v_{lab}/v_{cm})},  \label{eqlabangletoCM}
\end{equation}%
where%
\begin{equation}
\gamma =\frac{1}{\sqrt{1-(v_{lab}/c)^{2}}}.  \label{eqgammlab}
\end{equation}%
Here \cite{rpp_pdg_2022}%
\begin{equation}
\boldsymbol{\beta }_{cm}=\frac{\mathbf{p}_{lab}}{E_{lab}+M_{2}};\quad \text{and}%
\quad p_{cm}E_{cm}=p_{lab}M_{2}.  \label{eqbetacm}
\end{equation}%
Using eq. (\ref{eqlabangletoCM}) we obtain an expression for the cosine
derivatives%
\begin{equation}
\frac{d(\cos \theta _{cm})}{d(\cos \theta _{lab})}
=\frac{\gamma ^2 (v_{lab}/v_{cm}+\cos\theta _{cm})^3}{v_{lab}/v_{cm} \cos\theta _{cm}+1}\left(\frac{1-\cos\theta _{cm}^2}
{\gamma ^2 (v_{lab}/v_{cm}+\cos\theta _{cm})^2}+1\right)^{3/2},
\label{eqangletrans}
\end{equation}%
which can be used in eq.(\ref{eqDCSlabtoCM2}) to obtain the laboratory frame
differential cross section.

\subsection{The coupling constant}

we can estimate the value of the coupling constant $\lambda $ by using
information on the NN nuclear potential depth%
\begin{equation*}
V_{NN}=-38\text{ MeV}.
\end{equation*}%
At deuteron radius $r=2.1256$ fm \cite{hernandez2018}, the
Coulomb's potential is thus%
\begin{equation*}
V_{C}=\frac{e^{2}}{4\pi \epsilon _{0}r}=\frac{1\centerdot 44\text{ MeV.fm}}{%
2\centerdot 1256\text{ fm}}=0\centerdot 677449403\text{ MeV}.
\end{equation*}%
Therefore%
\begin{equation*}
\frac{\left\vert V_{NN}\right\vert }{V_{C}}=\frac{38\text{ MeV}}{0\centerdot
677449403\text{ MeV}}=56\centerdot 09275=\frac{\lambda }{\alpha }
\end{equation*}%
$\lambda $ is $56\centerdot 09275\alpha $ where $\alpha $ is the
electromagnetic coupling constant 
\begin{equation*}
\alpha =\frac{e^{2}}{4\pi \epsilon _{0}}/\hbar c=\frac{1}{137}
\end{equation*}%
Thus%
\begin{equation}
\lambda =\frac{56\centerdot 09275}{137}=0\centerdot 4094361314.
\label{eqcouplingconstant}
\end{equation}%
The fact that $\lambda <1$ makes the choice of the 1$^{st}$ order
interaction term, obtained from the expansion of the $T$-matrix, as a leading
contributor to the scattering cross section a reasonable approximation. Here
we have $\sigma \propto \lambda ^{4}=2\centerdot 81\times 10^{-2}$. The
second non-vanishing interaction term for the $T$-matrix includes two-pion
exchange. In this case we have $\sigma \propto \lambda ^{8}=7.9\times
10^{-4} $. The contribution is thus two orders of magnitude less than the
first order. This contribution is very minimal and we expect has negligible
impact in our results versus the experimental values.
\section{Results and disscussions}

\subsection{Elastic scattering}

\begin{figure}[ht]
\begin{center}
\includegraphics[scale=0.75]{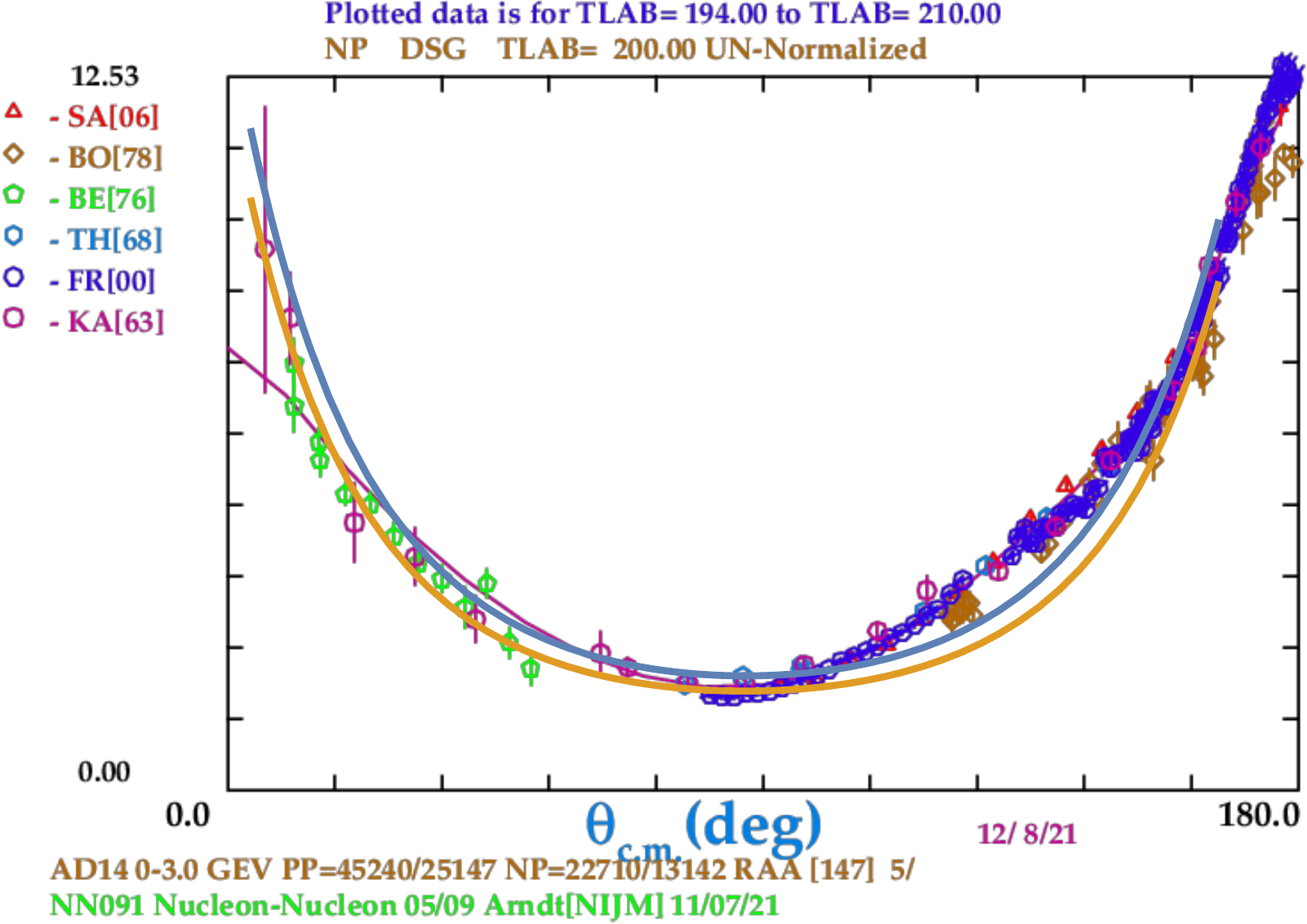}
\end{center}
\par\vspace{-0.7cm}
\caption{The calculated elastic proton-neutron scattering differential cross section using scalar current in eq. (\ref{eq14}) for $T_\mathrm{lab}=194\,\mathrm{MeV}$  ($E_{\mathrm{cm}}=1.974\,\mathrm{GeV} $) solid blue line and  $T_\mathrm{lab}=210\,\mathrm{MeV}$
($E_{\mathrm{cm}}=1.982\,\mathrm{GeV}$) solid orange line. The Calculated differential cross-section is compared with the experimental data taken from The SAID Database \cite{arndt1994,arndt1997,arndt2000,SAID}.}
\label{figDCS_SAID}
\end{figure}

Fig.(\ref{figDCS_SAID}) shows the calculated elastic proton-neutron (pn) scattering differential cross section using scalar current in eq. (\ref{eq14}) for  $T_\mathrm{lab}=194\,\mathrm{MeV}$  ($E_{\mathrm{cm}}=1.974\,\mathrm{GeV} $) solid blue line and  $T_\mathrm{lab}=210\,\mathrm{MeV}$
($E_{\mathrm{cm}}=1.982\,\mathrm{GeV}$) solid orange line. The Calculated differential cross-section is compared with the experimental data taken from The SAID Database \cite{arndt1994,arndt1997,arndt2000,SAID}. The experimental data is given on an arbitrary scale. Thus we have to fix the scales of the calculated differential cross-section to match those given in the experimental data, starting from zero for both axes. The general trend for both data is similar, however, the energy-dependent in the calculated data is stronger. The calculated data is symmetric about $\theta_{cm}=90^\circ$. For $\theta_{cm}>90^\circ$ the experimental data rise faster than the calculated one. Such a rise is the due to exchange effect, during the collision, the neutron and proton exchange places. That is, the forward-moving neutron becomes a proton and the backward-moving (in the center-of-mass system) proton becomes a neutron \cite{krane2008}.

\begin{figure}[ht]
\begin{center}
\includegraphics[scale=0.75]{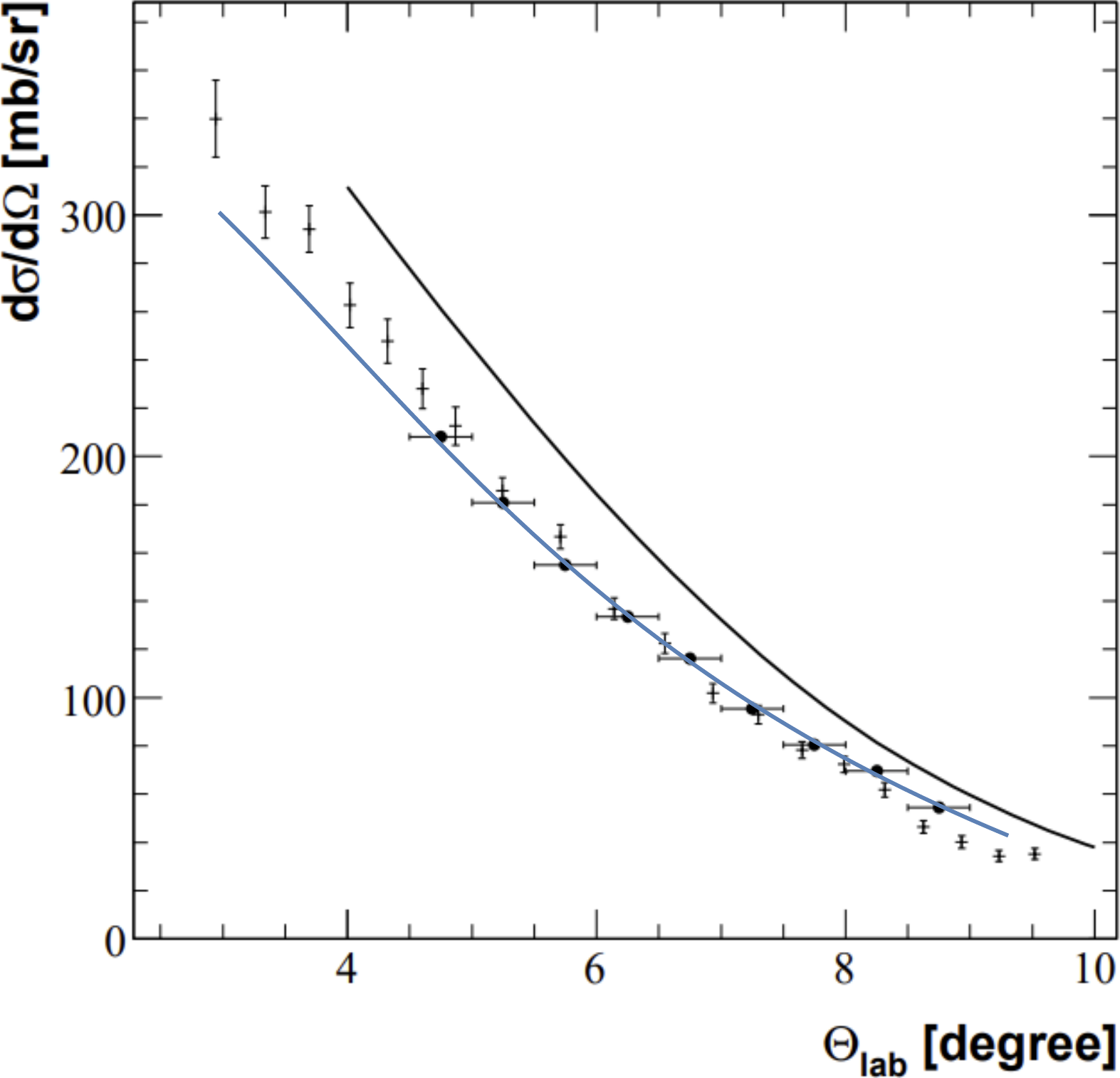}
\end{center}
\par\vspace{-0.7cm}
\caption{Laboratory frame elastic differential cross section as a function of the lab frame angle for pp scattering at 2.65 GeV claculated using scalar current (blue line). The calculated data is compared with the experimental values of reference \cite{chiladze2009}.}
\label{figDCS_pp}
\end{figure}

Fig.(\ref{figDCS_pp}) shows the laboratory frame elastic differential cross section as a function of the lab frame angle for pp scattering at 2.65 GeV calculated using scalar current (blue line). The calculated data is compared with the experimental values of reference \cite{chiladze2009}. The closed circles are the ANKE points \cite{chiladze2009}. The curve is the SP07 solution from the SAID analysis group \cite{arndt1994,arndt1997,arndt2000,SAID} and the crosses are experimental data at 2.83 GeV. The best fit is obtained for $\lambda=0.713$. The agreement is excellent. The fact that varying a multiplicative factor to get such an agreement indicates the success of the theory. We only need more data to confirm the value of $\lambda$.

\begin{figure}[ht]
\begin{center}
\includegraphics[scale=0.7]{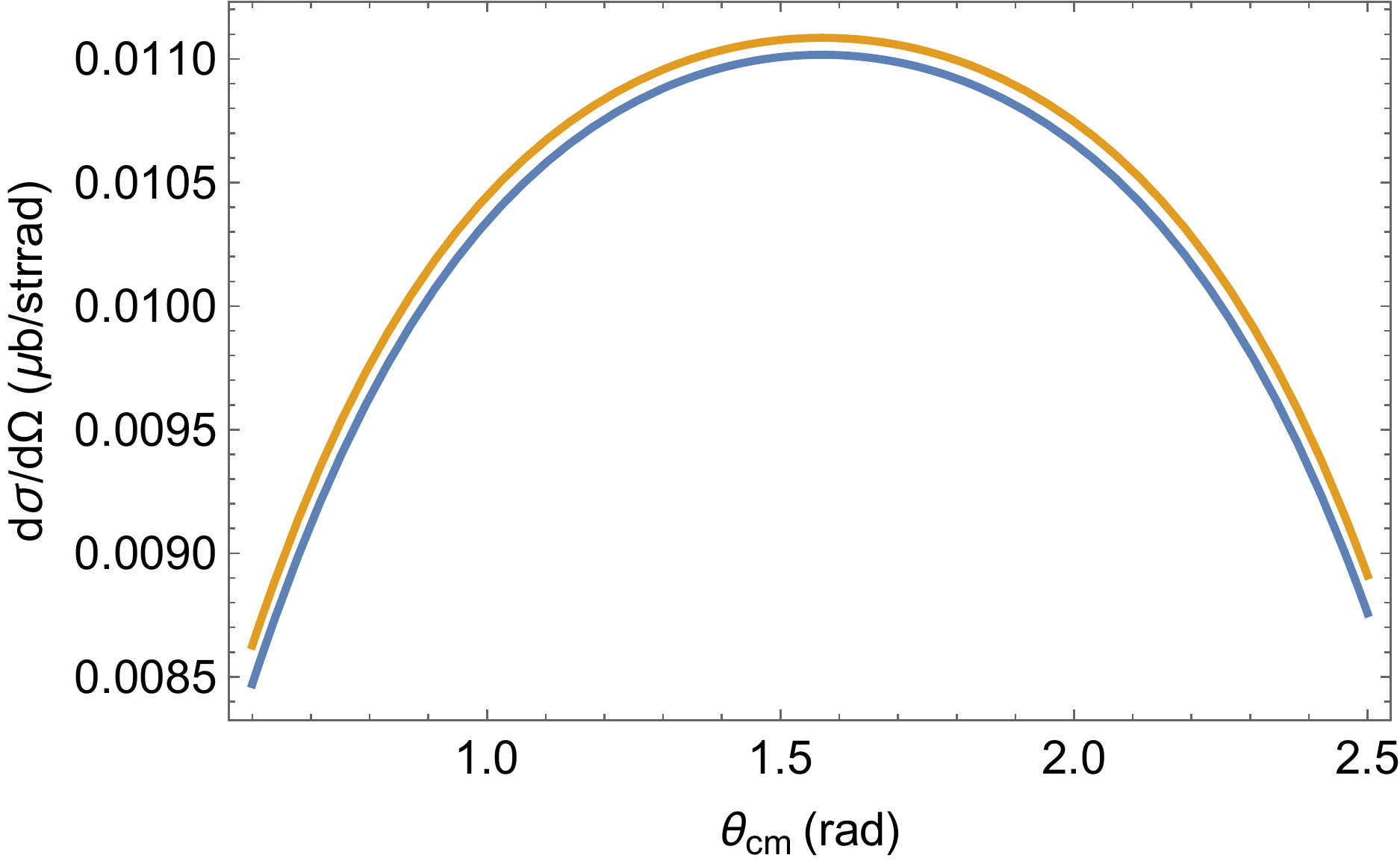}
\end{center}
\par\vspace{-0.7cm}
\caption{The NN elastic scattering differential cross section calculated using pseudoscalar current.}
\label{fig_eDCS_g5}
\end{figure}

Fig.(\ref{fig_eDCS_g5}) shows the NN elastic scattering differential cross section calculated using pseudoscalar current. The main feature is that the particles prefer to scatter along the transverse direction. The values of the differential cross section are six orders of magnitude smaller than those calculated using scalar current. This indicates that the contribution of the pseudoscalar current to interaction is negligible.

\begin{figure}[ht]
\begin{center}
\includegraphics[scale=1.0]{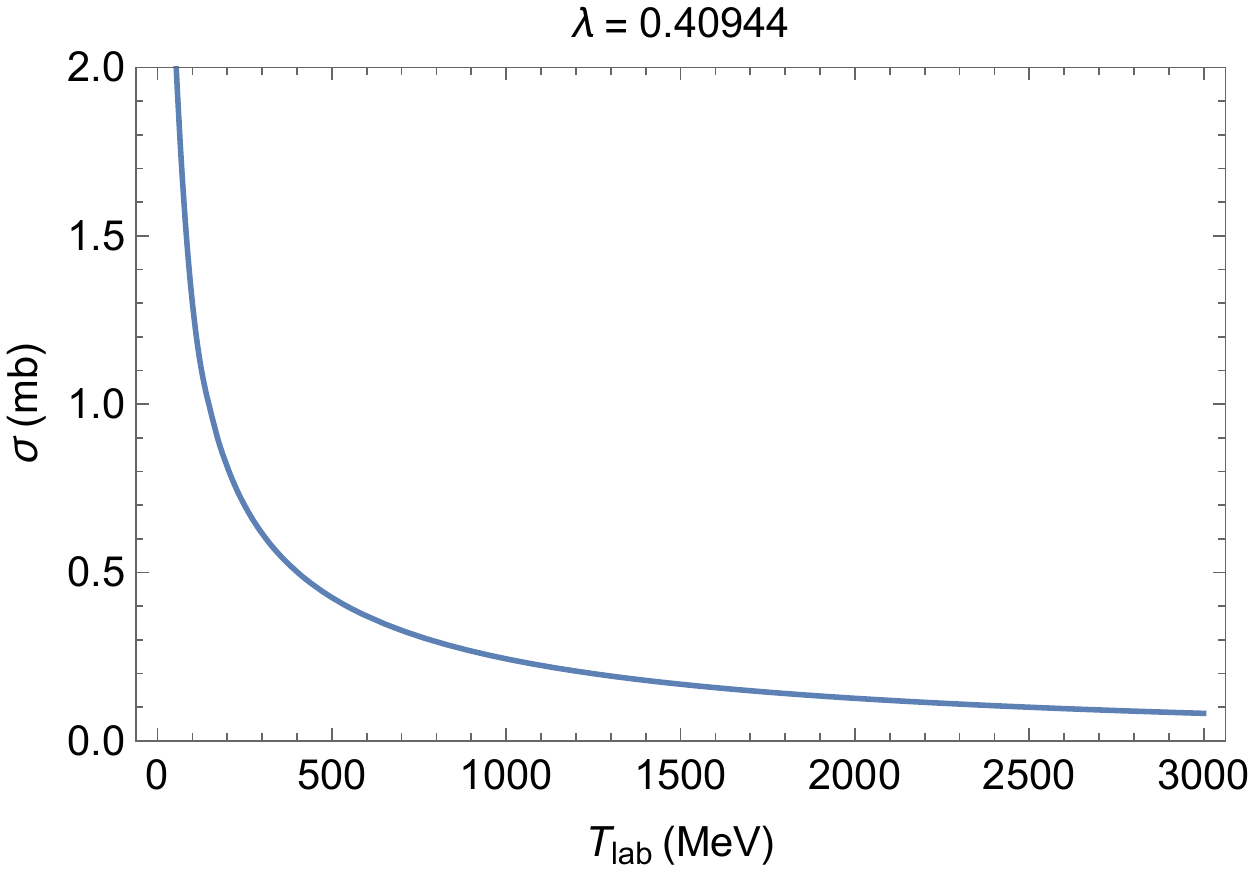}
\end{center}
\par\vspace{-0.7cm}
\caption{The NN elastic scattering cross section as a function of laboratory kinetic energy $T_{lab}$ in MeV.}
\label{fig_eCS}
\end{figure}

Fig.(\ref{fig_eCS}) shows the NN elastic scattering cross section as a function of laboratory kinetic energy $T_{lab}$ in MeV. As $T_{lab}$ increases cross section rapidly decreases. The result agrees with the result obtained in ref. \cite{norbury2010}. The rapid decrease in the elastic scattering cross section with increasing energy is because the inelastic scattering probability increases at the expense of the elastic scattering. The threshold kinetic energy for pion production is 279 MeV, and exceeding this threshold energy suppresses the probability of elastic scattering.

\begin{figure}[ht]
\begin{center}
\includegraphics[scale=1.0]{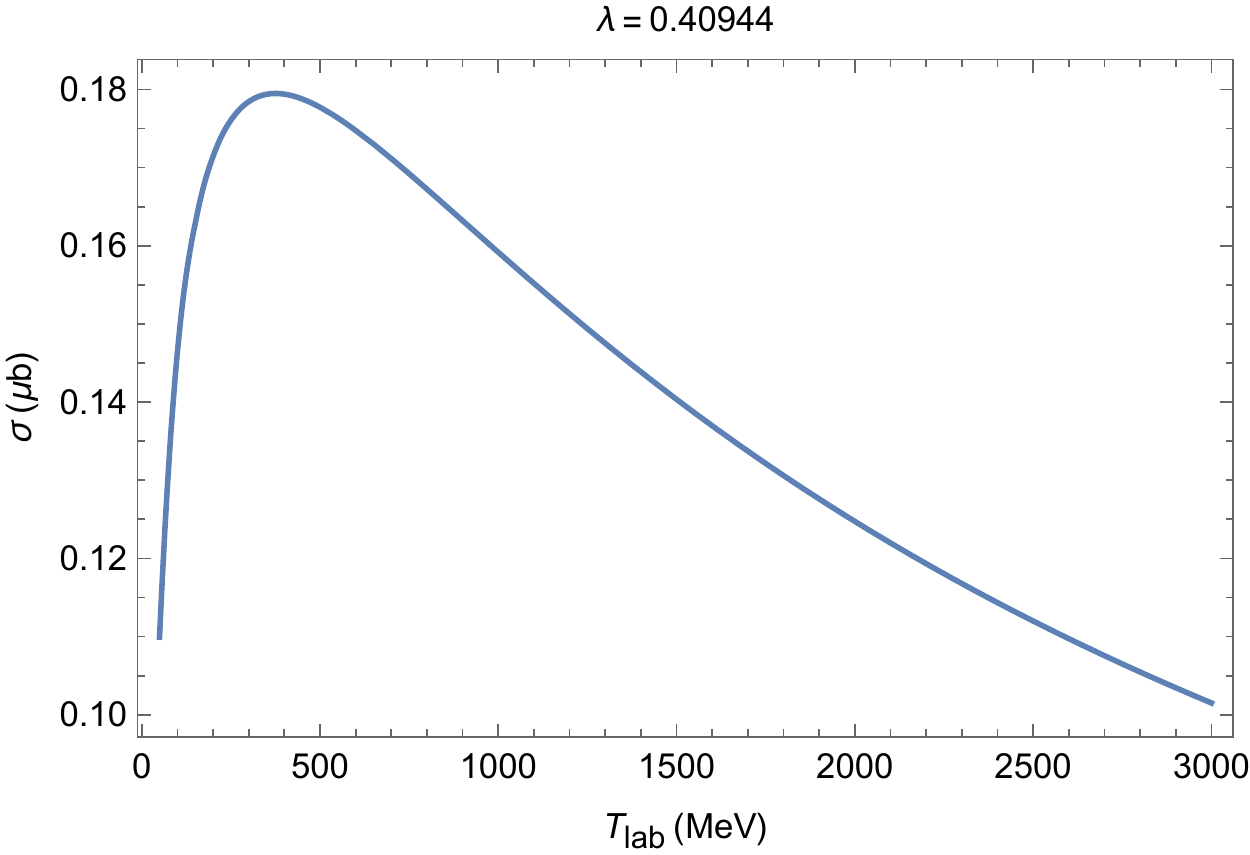}
\end{center}
\par\vspace{-0.7cm}
\caption{The NN elastic scattering cross section as a function of laboratory kinetic energy $T_{lab}$ in MeV. The cross-section is calculated using pseudoscalar current.}
\label{fig_eCSg5}
\end{figure}

Fig.(\ref{fig_eCSg5}) shows the NN elastic scattering cross section as a function of laboratory kinetic energy $T_{lab}$ in MeV. The cross-section is calculated using pseudoscalar current. In this case, the cross-section increases rapidly to a maximum value around $T_{lab}=300\,\mathrm{MeV}$, just above the pion production threshold. Afterward, the cross-section decreases as $T_{lab}$ increases. Again this is a sign of inelastic scattering contribution at the expense of elastic scattering. The cross-section calculated using pseudoscalar current is four orders of magnitudes smaller than that calculated using scalar current.

\subsection{Inelastic scattering}

To illustrate the inelastic scattering, we calculate the cross-section of production $\Delta(1232)$ via the reaction $p+p\to p+\Delta$. $\Delta$ has mass $M_{\Delta}=1232\,\mathrm{MeV}$, with spin $j^{\pi}=\frac32^{+}$ and isospin $t=\frac32$. In other words, the reaction is allowed by spin and isospin conservation laws.

\begin{figure}[ht]
\begin{center}
\includegraphics[scale=1.0]{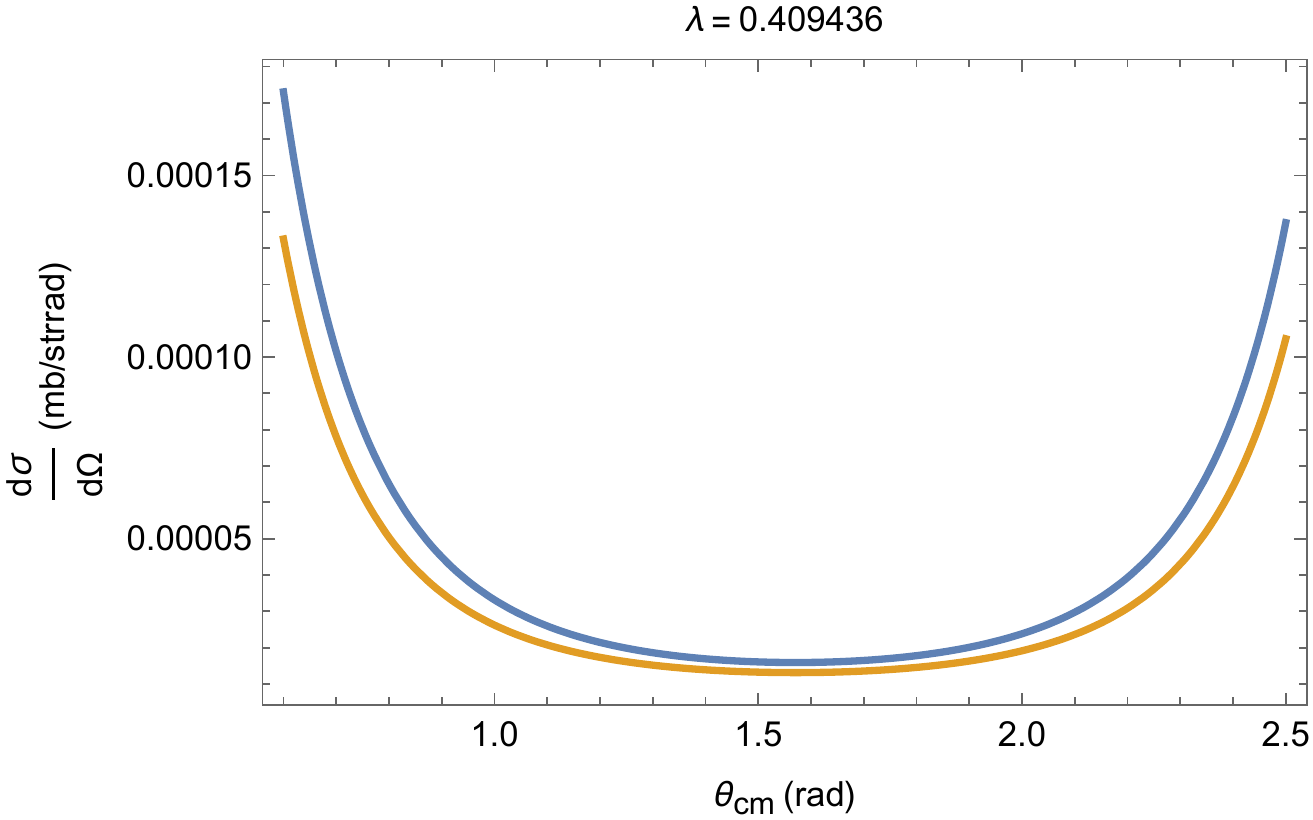}
\end{center}
\par\vspace{-0.7cm}
\caption{The differential cross section of $N+N\to N+\Delta(1232)$ production as a function of laboratory kinetic energy $T_{lab}$ in MeV. The cross-section is calculated using scalar current.}
\label{fig_iDCS}
\end{figure}

\begin{figure}[ht]
\begin{center}
\includegraphics[scale=1.0]{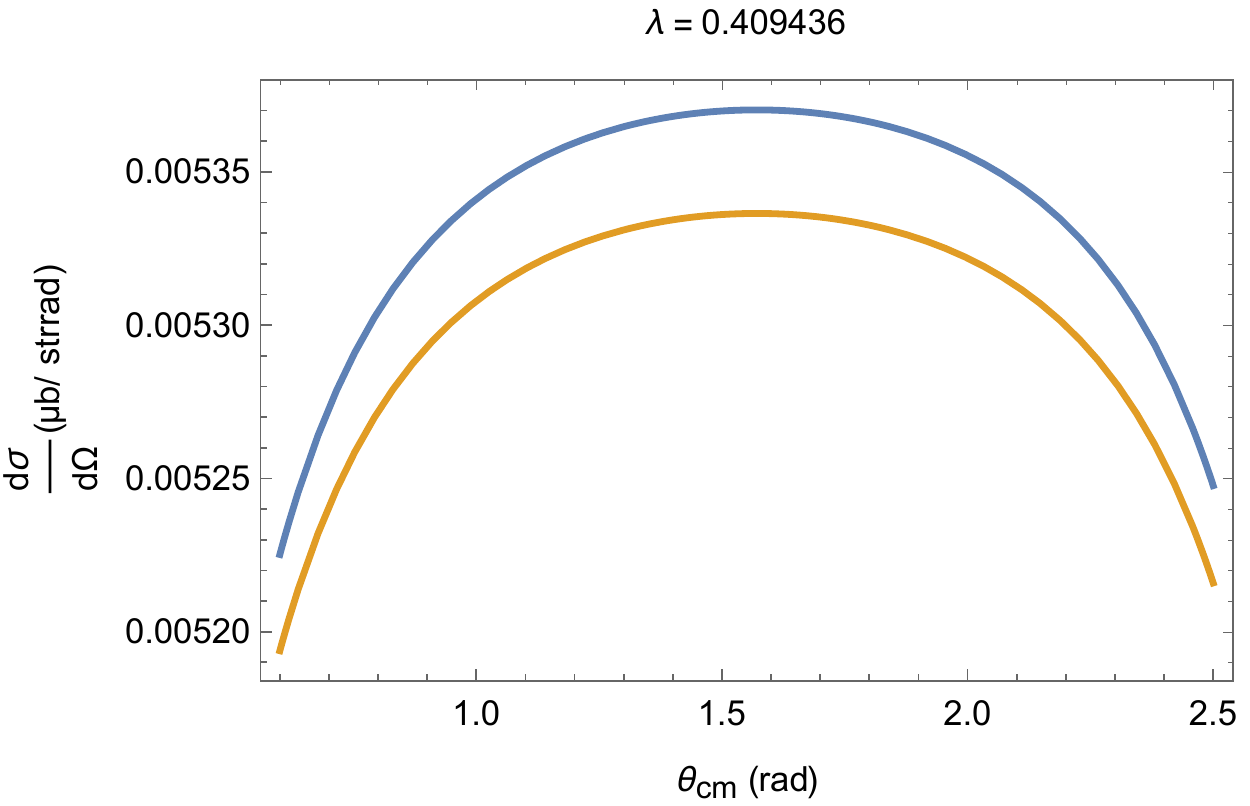}
\end{center}
\par\vspace{-0.7cm}
\caption{The differential cross section of $N+N\to N+\Delta(1232)$ production as a function of laboratory kinetic energy $T_{lab}$ in MeV. The cross-section is calculated using pseudoscalar current.}
\label{fig_iDCS_g5}
\end{figure}

Fig.(\ref{fig_iDCS}) shows the calculated inelastic $N+N\to N+\Delta(1232)$ scattering differential cross section using scalar current. The behavior is similar to that for the elastic differential cross section. Fig.(\ref{fig_iDCS_g5}) shows the $N+N\to N+\Delta(1232)$ inelastic scattering differential cross section calculated using pseudoscalar current. Again the transverse direction for scattering has the highest probability. The values of the differential cross section are slowly varying with $T_{lab}$ and they are three orders of magnitude smaller than those computed using scalar current.

\begin{figure}[ht]
\begin{center}
\includegraphics[scale=1.0]{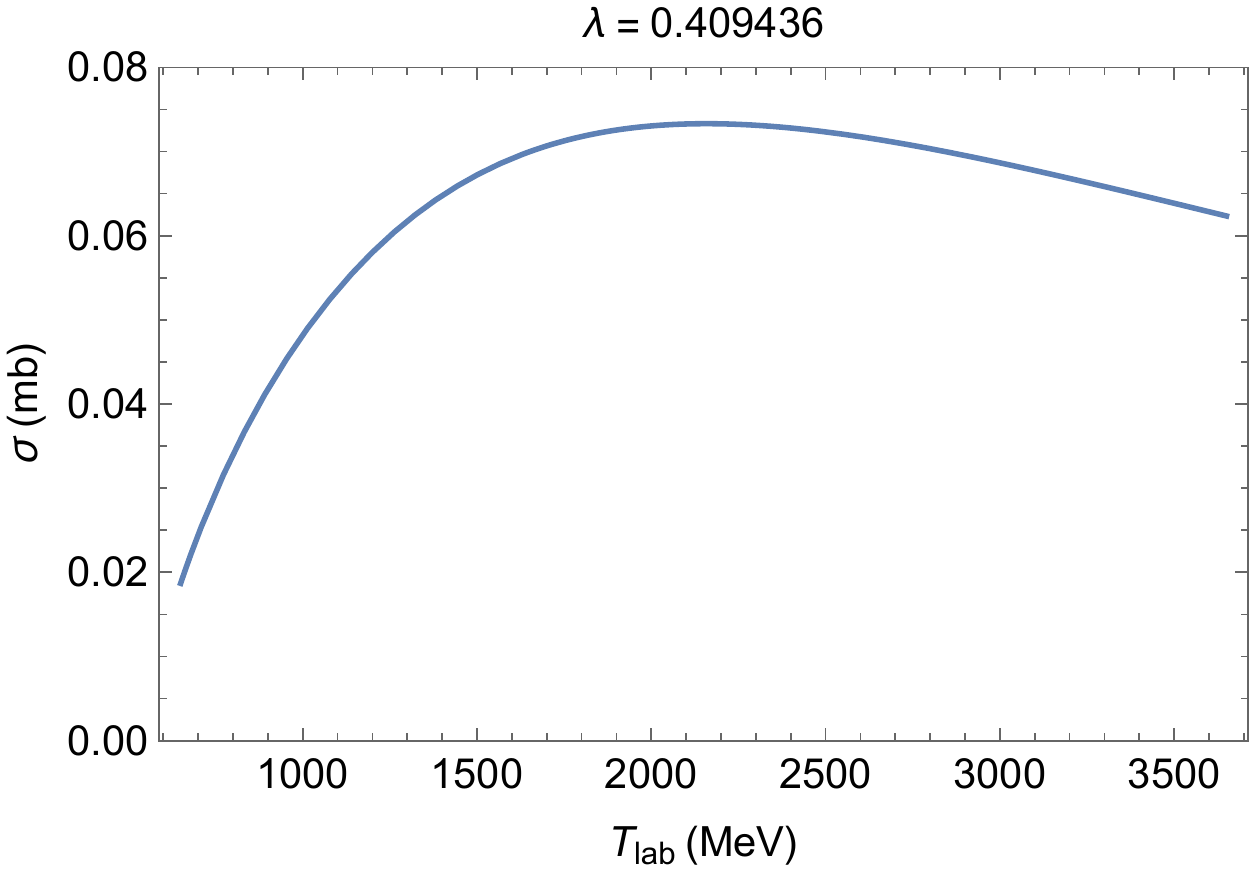}
\end{center}
\par\vspace{-0.7cm}
\caption{The inelastic $p+p\to p+\Delta(1232)$ cross section using scalar current, as a function of laboratory kinetic energy $T_{lab}$ in MeV.}
\label{fig_iCS}
\end{figure}

Fig.(\ref{fig_iCS}) shows the calculated inelastic $p+p\to p+\Delta(1232)$ cross section using scalar current. The trend is similar, however lower, to the calculated data in fig.8 of ref.\cite{norbury2010}. Again we need to vary the coupling constant $\lambda$ to achieve the best fit for the experimental data. We couldn't overlay our curve on fig.8 of ref.\cite{norbury2010} because of the logarithmic scale. The cross-section reaches maximum at $T_{lab}=2.00\,\mathrm{GeV}$, saturates, and then decreases slowly when $T_{lab}>2.30\,\mathrm{GeV}$.

\begin{figure}[ht]
\begin{center}
\includegraphics[scale=1.0]{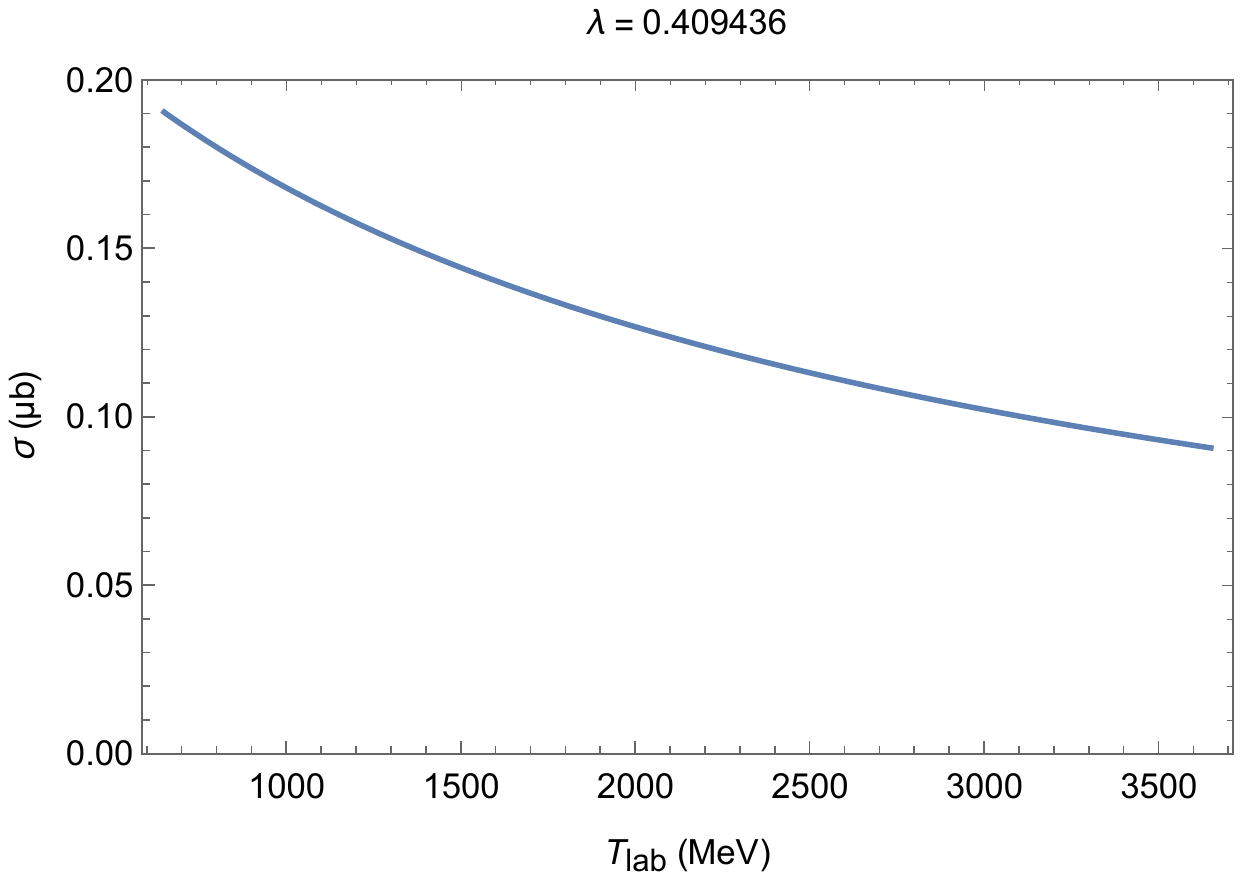}
\end{center}
\par\vspace{-0.7cm}
\caption{The inelastic $N+N\to N+\Delta(1232)$ cross section as a function of laboratory kinetic energy $T_{lab}$ in MeV. The cross-section is calculated using pseudoscalar current.}
\label{fig_iCS_g5}
\end{figure}

Fig.(\ref{fig_iCS_g5}) shows the calculated $p+p\to p+\Delta(1232)$ cross section using psuedoscalar current. The curve doesn't reflect the expected behavior for the inelastic cross section. The values of the cross-section are two orders of magnitudes smaller than those calculated using scalar current. 

\subsection{Conclusion}
The success of any scientific theory is measured by its ability to accurately explain and predict observations and experimental results. 
The pion exchange theory, using the interaction of the fermion field with the scalar field can describe experimental cross-section data.

One of the most important aspects of the theory is its ability to predict the behavior of NN scatterings based on a real physical model that describes the underlying mechanisms of the interaction. This gives confidence in the theory and encourages us to use it for further investigations and testing. This is important to obtain solid value(s) of the coupling constant $\lambda$. It is expected that $\lambda$ may vary with momentum transfer, thus it can be a function of the scattering energy.

The pseudoscalar current model, though fundamentally sounds accurate, it does not offer a reliable description of the experimental results. In any case, its contribution is negligible and we advise excluding it from any future calculations. 

The simplicity of the theory makes it all the more impressive. The most powerful scientific theories are often those that can describe complex phenomena with simple equations or models. By making this theory easily understandable and replicable, enables others to build upon this work.

\clearpage
\pagebreak
\bibliographystyle{apsrev4-1}
\bibliography{Content/reference}

\end{document}